\documentclass[preprint,12pt]{elsarticle}

\usepackage{natbib}
\usepackage{geometry}
\usepackage{fleqn}
\usepackage{graphicx}
\usepackage{newtxtext,newtxmath}
\usepackage{hyperref}
\usepackage{xurl}
\usepackage{amsmath}

\usepackage{amssymb}

\usepackage{float}
\usepackage[version=4]{mhchem}
\newenvironment{DIFnomarkup}{}{}

\hypersetup{
    colorlinks,
    linkcolor={red!50!black},
    citecolor={blue!50!black},
    urlcolor ={blue!80!black},
    pdfauthor={Rethabile Kgolobe}
}

\usepackage{booktabs}
\usepackage[justification=centering]{caption}

\usepackage{framed}  
\usepackage{nomencl} 
\makenomenclature

\usepackage{color,soul}
\usepackage{subcaption}

\usepackage{bm}

\usepackage[detect-all]{siunitx} 

\graphicspath{{./figs/}}
\journal{Annals of Nuclear Energy}

\usepackage{multicol}
\renewcommand*{\nompreamble}{\begin{multicols}{2}}
\renewcommand*{\nompostamble}{\end{multicols}}
\usepackage{totcount}
\newtotcounter{TODO}

\usepackage{marginnote}

\begin{document}

\begin{frontmatter}

\title{Unsupervised Anomaly Detection in Measured Neutron Flux Profiles of the SAFARI-1 Research Reactor}

\author[EEUJ,NECSA]{Rethabile Kgolobe\corref{correspondingauthor}}
\ead{rethabile.kgolobe@necsa.co.za}
\author[NECSA]{Pavel M. Bokov}
\author[MEUJ]{Simon H. Connell}
\author[EEUJ]{Charis Harley}
\author[MEUJ,NECSA]{Bongani G. Maqabuka}
\author[NECSA]{Lesego E. Moloko}

\cortext[correspondingauthor]{Corresponding author}

\address[EEUJ]{Electrical and Electronic Engineering Science, University of Johannesburg, Johannesburg,
South Africa}
\address[MEUJ]{Mechanical Engineering Science, University of Johannesburg, Johannesburg, South Africa}
\address[NECSA]{Research and Innovation, The South African Nuclear Energy Corporation SOC Ltd (Necsa), Building-1900, P.O. Box 582, Pretoria 0001, South Africa}

\begin{abstract}
Knowledge of the neutron flux distribution is essential for safe and effective reactor utilisation. In the SAFARI-1 research reactor, a set of axial flux profiles are periodically measured by activating natural copper wires inserted along the height of the fuel or control assemblies and subsequently measuring their activity using a sodium iodide detector. Improper wire insertion can, however, result in profiles that do not accurately represent the actual neutron flux distribution. Building on previously identified potential indicators of improper wire insertion, an automated approach based on unsupervised learning was developed to detect anomalous axial flux measurements. The measured data were preprocessed using missing-value imputation, decay correction, and normalisation. High-order polynomial fitting was then applied to characterise the axial flux profiles and derive features describing their shape. The Density-Based Spatial Clustering of Applications with Noise (DBSCAN) algorithm was employed to group flux profiles into regular and abnormal clusters based on these features with results matching the known groups. In addition, Convolutional Autoencoders were developed and trained on flux profiles from central and peripheral core positions to identify deviations from the regular profiles through reconstruction error, achieving better performance in central assemblies than in peripheral assemblies. The developed framework is intended to provide feedback to reactor operation by identifying potential wire-insertion problems in the manual technique used for axial flux monitoring at SAFARI-1.
\end{abstract}


\end{frontmatter}
\begin{DIFnomarkup}
\begin{table*}[!t]
    \begin{framed}
        \section*{Nomenclature}
        \begin{multicols}{2}
            \begin{description}
                \itemsep-0.1em\small
                \item[\rm ANN]      Artificial Neural Network
                \item[\rm BNN] 		Bayesian Neural Network
                 \item[\rm CAE]		Convolutional Autoencoder
                 \item[\rm DBSCAN]  Density-Based Spatial Clustering of Applications with Noise
                \item[\rm DL]       Deep Learning
                \item[\rm DNN]      Deep Neural Network
                \item[\rm DT]       Decision Tree
                \item[\rm GS]       Grid Search
                \item[\rm IAEA]     International Atomic Energy Agency
                \item[\rm MAE]      Mean absolute error
                \item[\rm MTR]      Material Testing Reactor
                \item[\rm MCD]      Monte Carlo Dropout
                \item[\rm ML]       Machine Learning
                \item[\rm MAE]      Mean Absolute Error
                \item[\rm RS]       Random Search
                \item[\rm SAFARI-1]	South African Fundamental Atomic Research Installation, generation 1
                \item[\rm SVM]      Support Vector Machine
            \end{description}
        \end{multicols}
        \vspace{-2mm}
    \end{framed}
\end{table*}
\end{DIFnomarkup}

\section{Introduction}\label{Introduction}

Nuclear reactor safety requires a reliable assessment of the reactor core state to ensure that operating conditions remain within established safety limits and that deviations from expected performance can be identified promptly \cite{IAEA_TECDOC_1332_2003}. A key quantity for assessing the reactor core state is the neutron flux distribution, which characterises the spatial and temporal variation of the neutron population within the core \cite{Lamarsh2018reactortheory}. Knowledge of this distribution provides a basis for determining important reactor parameters, including core power and the location of regions with elevated power density and consequently higher fuel cladding temperatures. It can also support the identification of deviations from the expected core state, such as fuel misloads, as well as inconsistencies in the measurement system, such as axially shifted flux measurements \cite{Kgolobe2025SAIP}.

In research reactors, activation measurements provide a means of assessing aspects of the reactor core state, particularly the neutron flux distribution. Activation foils and wires, including copper wires, are widely used for this purpose. The resulting activation data can be used to characterise the spatial distribution of neutron flux, validate neutronic calculations, infer power distributions, and identify changes in the reactor core state over time \cite{Chapman1965determination, Musa2012determination}. Copper-wire activation measurements are performed by inserting copper wires into the reactor core at selected locations, irradiating them for a prescribed period, measuring the induced activity along the wires after irradiation, and analysing the resulting activity profiles to reconstruct the axial neutron flux distribution at each insertion location \cite{Moloko2021estimation}. However, procedural inconsistencies during wire insertion and handling can introduce measurement artefacts that might resemble genuine physical anomalies, making it difficult to differentiate measurement error from an actual anomaly in the core \cite{Moloko2021estimation}.

In practice, the flux measurement procedure still relies on manual, expert-driven inspection or comparison against simulation code predictions, both of which are time-consuming, subjective, and difficult to scale across many cycles and assemblies \cite{Chapman1965determination, Moloko2022quantification}. Anomaly detection in this monitoring process, therefore, presents an opportunity to apply a Machine Learning (ML) algorithm. However, some ML algorithms, such as supervised learning techniques, require labelled anomaly examples, which do not exist for these measurements. In contrast, unsupervised learning techniques such as Density-Based Spatial Clustering of Applications with Noise (DBSCAN) and Convolutional Autoencoders (CAEs) can learn the characteristics of normal flux distribution from historical data and flag deviations without requiring labelled anomalies \cite{articleFiase, KOLLIAS2022109293}.

The application of ML to nuclear reactor problems has yielded promising results, ranging from parameter estimation to anomaly detection \cite{ioannou2021intelligent, chaudhary2024anomaly}. For example, Garis et al. \cite{garis1998determination} aimed to determine the axial position of control rods by unfolding them from the asymmetric axial neutron flux profiles caused by partial rod insertion. To address this issue, the authors used a so-called three-layered neural network trained via backpropagation on flux profiles generated by the Studsvik core master system \cite{Huang01081994}. Testing demonstrated successful unfolding, yielding root-mean-squared errors ranging from $\SIrange{-3.3}{3.3}{\centi\meter}$ and standard deviation ranging from $\SIrange{0.63}{1.91}{\centi\meter}$, with and without Gaussian noise.

Terman et al. \cite{terman2018determination} reconstructed critical reactor states, specifically the axial positions of five control rods and the fuel burn-up level, for the Tehran research reactor using two-dimensional neutron flux measurements. The authors employed multi-layered feed-forward Artificial Neural Networks (ANNs) trained with a Bayesian regularisation backpropagation algorithm. Their model achieved excellent state reconstruction capabilities, with a standard deviation of $\SI{2.15E-3}{\percent}$ and a mean of $\SI{-1.62E-5}{\percent}$, minimising estimation errors and demonstrating robust generalisation.

In the context of research reactors, Moloko et al. \cite{Moloko2021estimation, Moloko2022quantification} aimed to approximate axial neutron flux measurements and quantify the prediction uncertainties of neural networks applied to SAFARI-1 axial neutron flux data. The researchers developed three deep learning approximations: Deep Neural Networks (DNNs), Monte Carlo Dropout (MCD), and Bayesian Neural Networks (BNNs), solved via variational inference. Testing on unseen cycle data indicated that all three architectures accurately predict axial neutron flux profiles with strong generalisation capabilities and highlighted that the applied uncertainty quantification enhanced the credibility of the models. The results yielded normalised root-mean-square errors of about $\SIrange{7}{9}{\percent}$ for fuel assemblies and up to $\SI{15}{\percent}$ for follower assemblies.

Among ANN architectures, an autoencoder is an unsupervised learning approach consisting of an encoder and a decoder. Autoencoders are used for a range of tasks, including image processing, anomaly detection, and generative modelling. \cite{mienye2025deep}. Autoencoders learn patterns in the data using an encoder that maps the input to the latent space, and a decoder that maps the latent representation back to the original input \cite{sewak2020overview}. This approach has been utilised in a variety of applications, such as the medical field \cite{nawaz2024ensemble}, cybersecurity \cite{Senthilcyber}, and finance \cite{articleDaoud}. In the context of anomaly detection in nuclear reactors, Sun et al. \cite{sun2024unsupervised} aimed to integrate convolutional adversarial autoencoders with a self-attention mechanism to detect anomalies in multivariate time-series data from a nuclear power plant. The results obtained a higher $F_1$ score of $\SI{100}{\percent}$ without noise and $\SI{94.58}{\percent}$ with noise included, outperforming other unsupervised learning methods, identifying the anomalous operation states within $\SI{0.04}{\second}$. These findings provide a solid foundation for using unsupervised CAEs to process data subject to high operational statistical noise; the proposed method was limited to simulated data rather than physical experimental measurements.

In the context of ML- and Deep Learning (DL)-driven anomaly detection and flux prediction in nuclear reactors, specifically when hyperparameter optimisation is applied, Racheal et al. \cite{racheal2022evaluation} evaluated the performance of ML models for nuclear reactor accident prediction using an augmented dataset that includes random noise to simulate sensor failures. The authors developed and compared three supervised models: Support Vector Machines (SVMs), Decision Trees (DTs), and Multilayer Perceptrons (MLPs), optimised using a grid search (GS). The results showed that hyperparameter tuning yielded the best parameter classification accuracies of $\SI{0.972}{\percent}$, $\SI{0.988}{\percent}$, and $\SI{0.988}{\percent}$ for SVMs, DTs, and MLPs, respectively, with the optimised SVMs and DTs models outperforming the MLP on noise-augmented inputs. This systematic tuning highlights how parameter adjustments improve an algorithm's performance in the presence of background noise. However, the study is restricted to a supervised learning classification framework.

Extending DL applications in the nuclear reactor domain, specifically in the SAFARI-1 research reactor, Moloko et al. \cite{Moloko2023prediction} developed an approach to predict axial neutron flux measurements using quantified uncertainties. The study employed DNNs and quantified model uncertainties using two methods: MCD and Bayesian Neural Networks solved via variational inference (BNNs VI). These models were optimised using random search (RS) and GS to find the best-performing parameters while minimising mean absolute error. The tuned models achieved coefficients of determination ($R^2$) values of 0.903 and 0.984 for the peripheral and radially central regions, respectively. These results provide insights into a framework for using hyperparameter tuning to optimise the performance of DL algorithms under high noise-to-signal ratios in physical copper-wire activation measurements.

Previous work by Kgolobe et al. \cite{Kgolobe2025SAIP} used statistical methods to investigate the relationship between shape features of the axial flux and the control bank positions. Specifically, the axial neutron flux distributions exhibit three characteristic extrema: a maximum point located near the centre of the profile, and a minimum point near the bottom and top ends of the flux profiles, as will be discussed in Subsection~\ref{Characteristics_of_Measured_Flux_Profiles}. The study indicated that the maximum points are more sensitive to control bank positions than the bottom and top minimum points. In particular, the top minimum points remained sufficiently stable to serve as reliable indicators of anomalous flux profiles caused by improperly inserted copper wires. Although this study demonstrated the value of intrinsic physical characteristics of flux profiles in identifying such anomalies, it remained limited to statistical characterisation. It did not extend to the development of an automated, data-driven anomaly detection system. Given the issues that can result from improperly inserted copper wires, this research aims to employ both DBSCAN and convolutional autoencoders (CAEs) to detect abnormal flux profiles. By applying these algorithms, the study seeks to automate anomaly detection, thereby providing feedback to improve flux measurement procedures and support safe and effective reactor operation.

The remainder of this paper is structured as follows. In Section~\ref{Problem_Definition}, the copper wire neutron flux measurement procedure and characteristics of the measured flux profiles are described. In Section~\ref{DataPreprocessing&Feature Extraction}, the data preprocessing techniques used in this study and the feature extraction techniques employed are presented. The methodology for applying the DBSCAN clustering algorithm to group relevant flux features is briefly summarised in Section~\ref{DBSCANForAnomalyDetection}. The training of the CAEs for anomaly detection, including the hyperparameter tuning technique employed, is presented in Section~\ref{ConvolutionalAutoencodersForAnomalyDetection}. The results and discussions are presented in Section~\ref{Results&Discussion}, followed by the summary of the findings and recommendations in  Section~\ref{Conclusiona&Recommendations}.

\section{Problem Definition}\label{Problem_Definition}

\subsection{SAFARI-1 Research Reactor}

\begin{figure}
    \centering
        \centering
        \includegraphics[width=\linewidth]{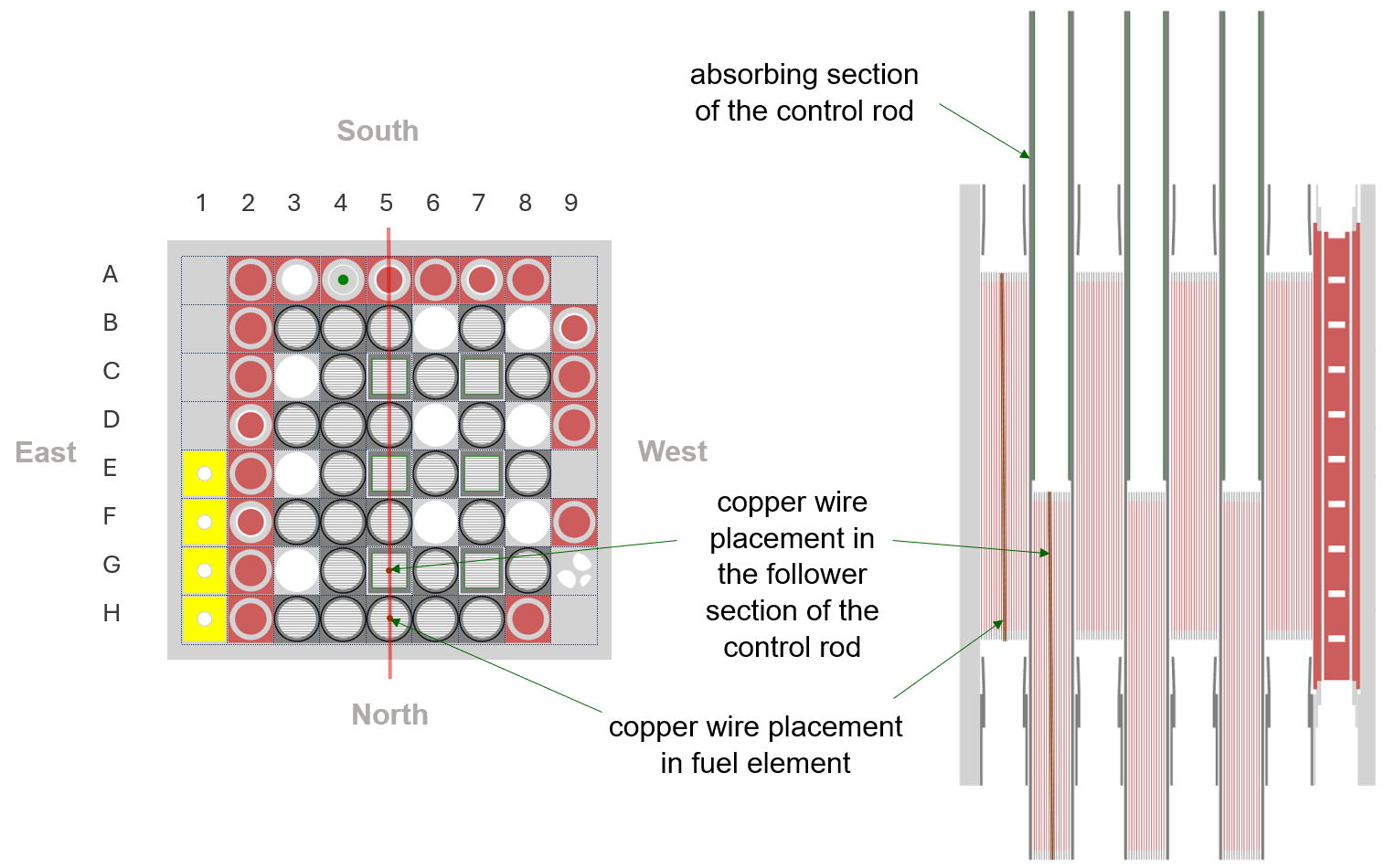}
    \caption{The OSCAR-5 \cite{Prinsloo2017recent} model for the SAFARI-1 research reactor -- the top view on the left and the vertical cross-section of the on the right with indicated placement of copper wire during the measurement.}
    \label{fig:SAFARI-1-top-view}
\end{figure}

The SAFARI-1 (South Africa Fundamental Atomic Research Installation, generation 1) is a pool-type Material Testing Reactor (MTR) operated by South African Nuclear Energy Corporation SOC Ltd (Necsa) located in Pelindaba, South Africa \cite{IAEA2022}. SAFARI-1 currently operates at $\SI{20}{\mega\watt}$ and is designed with an $8\times9$ core grid structure to accommodate 26 fuel assemblies and six control rod assemblies as shown in Figure~\ref{fig:SAFARI-1-top-view}. The control rod assemblies consist of the fuel follower and absorber part connected by an aluminium coupling piece \cite{Moloko2023prediction}. The fuel and follower assemblies contain $\SI{19.75}{\percent}$ enriched Uranium Silicide $\ce{U_{3}Si_{2}Al}$ with 19 and 15 fuel plates, respectively \cite{IAEA2022}. Additionally, the SAFARI-1 core is reflected by plugged, solid and hollow beryllium on the east, south and west sides of the core, with the north side facing the poolside facility \cite{Moloko2023prediction}. The core further consists of aluminium water boxes, ringas sample irradiation systems, assembly lead, a hydraulic rabbit for isotope production, and aluminium lead for structural support \cite{Moloko2023prediction}. The reactor is also used for neutron experiments, such as neutron radiography, with its beam lines running radially from the reactor core box.

The SAFARI-1 research reactor operates for 21–30 days per cycle, with a 5-day refuelling shutdown per cycle and a 12-day refuelling and maintenance shutdown per year \cite{Moloko2022quantification}. At the beginning of each operational cycle, neutron flux measurements are conducted in the SAFARI-1 when the reactor is operating at low power levels, about $\SI{200}{\kilo\watt}$ or $\SI{1}{\percent}$ of full power. The process of measuring neutron flux at SAFARI-1 is referred to as copper wire activation, in which natural copper wires within hollow aluminium blades are manually inserted into the 32 assemblies containing fissile material and extend beyond the height of the reactor core.
The copper wires are irradiated for a specified period, during which neutron-induced nuclear reactions produce radioactive nuclides within the wires.
This activation produces mainly two radioactive nuclides, namely: $\ce{^{66}Cu^{*}}$ and $\ce{^{64}Cu^{*}}$ with half lives of $T_{1/2} = \SI{5.1}{\minute}$ and $T_{1/2} = \SI{12.7}{\hour}$, respectively. After irradiation, the wires are removed from the assemblies within the core and are allowed to decay for some time to eliminate the activity of the $\ce{^66{Cu}^{*}}$ isotope \cite{Chapman1965determination}. The measurements of the activated copper wires are then performed using an assigned scanning system, equipped with a Sodium Iodide (NaI) detector, from which counts from $\ce{^{64}Cu^{*}}$ are measured at 180 equidistant points along the wires for all 32 assemblies (26 fuel plus six follower assemblies).

These measurements, obtained through copper wire activation, are crucial for determining the neutron flux distribution within the reactor core. The neutron flux measurements used in this study comprise data from 92 operational cycles from 2009 to 2024, along with the corresponding control rod bank positions. The data was sourced from both the International Atomic Energy Agency (IAEA) research benchmarking database \cite{IAEA2022} and Necsa. The central problem addressed in this study stems from the manual nature of the process in which copper wires are inserted into fuel-containing assemblies. Incorrectly inserted copper wires can prevent the wire from properly measuring the full flux distribution along the height of the fuel and follower assemblies, resulting in inaccurate neutron flux measurements and introducing anomalies into the data. Since decisions regarding reactor operation may be based on these measurements, such anomalies, if undetected, could ultimately affect reactor performance and core safety parameters.

\subsection{Characteristics of Measured Flux Profiles}\label{Characteristics_of_Measured_Flux_Profiles}

\begin{figure}
    \centering
    \begin{minipage}[b]{0.49\textwidth}
        \centering
        \includegraphics[width=\linewidth]{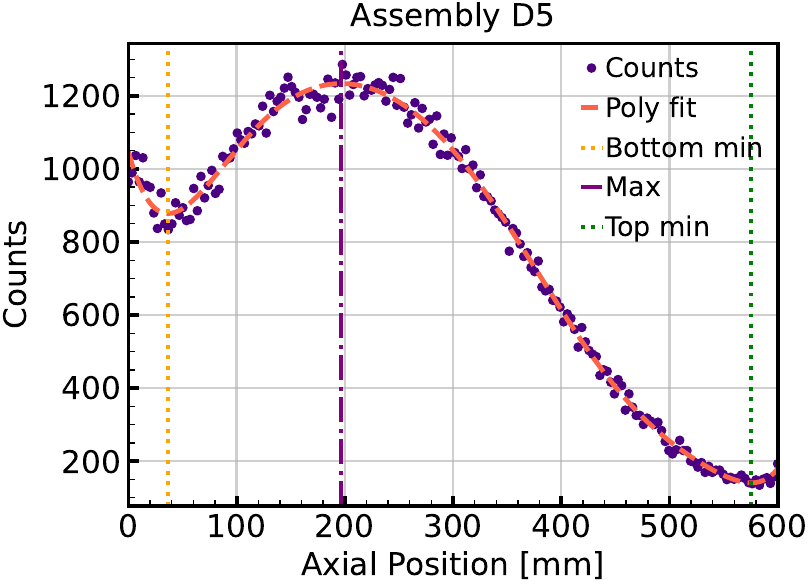}
    \end{minipage}%
    \begin{minipage}[b]{0.49\textwidth}
        \centering
        \includegraphics[width=\linewidth]{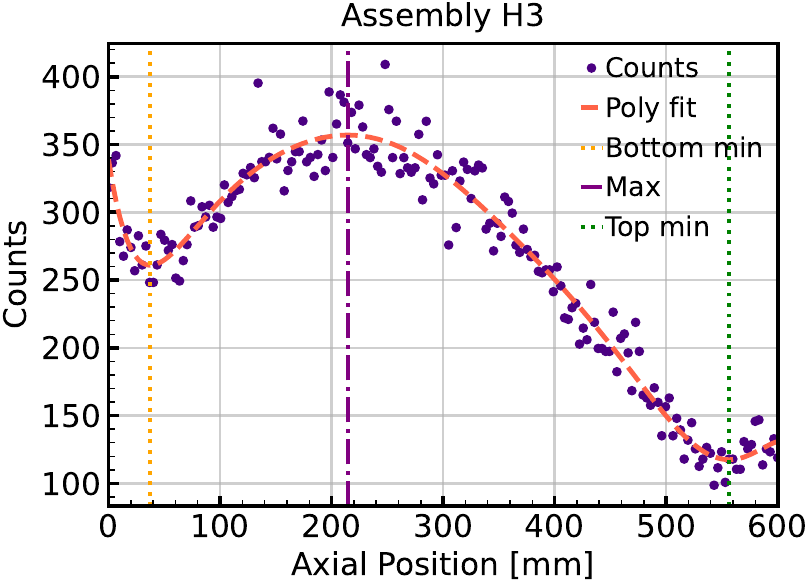}
    \end{minipage}
    \caption{Neutron flux measurements with flux extrema for fuel assemblies D5 (left) and H3 (right). In the legend, "Poly fit," "Bottom min," "Max," and "Top min" denote the polynomial fit, and the bottom minimum, maximum, and top minimum points of the flux profile, respectively.}
    \label{fig:fluxshape}
\end{figure}

The measured neutron (thermal) flux distributions of fuel assemblies exhibit three characteristic extrema: a maximum point and minimum points at the bottom and top ends of the profile, as illustrated in Figure~\ref{fig:fluxshape}. This truncated cosine axial shape results from the combined effects of neutron production, moderation, leakage, and absorption within the core. The central maximum corresponds to the regions of both the highest fission rate and the highest neutron flux. At the same time, the flux decreases towards the edges of the core due to neutron leakage and absorption by the surrounding water. Within this flux shape, secondary peaks occur near the top and bottom ends of the fuel assemblies as a result of fast neutrons being moderated by the adjacent water, thereby increasing the local thermal neutron flux. The flux increases slightly toward the axial boundaries as a result, yielding two local minima in the vicinity of the top and bottom boundaries of the active core. These axial minima and maxima represent, therefore, intrinsic physical characteristics observed in the thermal neutron flux measurements.

Later in this study, these pertinent physical characteristics of the neutron flux distribution are extracted as input features for the Machine learning (ML) algorithms to detect anomalies.
Hence, this approach combines reactor-physics-based features, namely the axial extrema and skewness of the neutron flux, with data-driven clustering and autoencoder-based algorithms for anomaly detection.

For control follower assemblies, which are displaced downward at the beginning of the operational cycle when the measurements are performed (see Figure~\ref{fig:SAFARI-1-top-view}), the flux shape is additionally influenced by the absorbing section above, moderation in the coupling piece, and the axial position of the follower relative to the active core. Consequently, their flux-shape features differ from those of fuel assemblies and may not exhibit the same behaviour with respect to anomalous measurements. The present study therefore focuses on fuel assemblies, while control follower assemblies are left for future investigation.

\section{Data Preprocessing and Feature Extraction}\label{DataPreprocessing&Feature Extraction}

Data preprocessing and feature extraction techniques were applied to the copper wire activation measurement data discussed in Section \ref{Introduction}. This involved missing-value imputation, decay correction, and normalisation, following which the axial flux features were extracted using high-order polynomials, and the skewness of the flux shapes was subsequently extracted. These tasks were performed using open-source Python libraries, namely \texttt{Pandas, Numpy, Scipy}, and \texttt{Sklearn}. Specifically, \texttt{Numpy} provided tools for array handling and numerical computing, while \texttt{Pandas} was used for data analysis and manipulation \cite{Pandas, bressert2012scipy}. \texttt{Scikit-learn} was utilised for data preprocessing, including normalisation and clustering, while \texttt{Scipy}, built on \texttt{Numpy}, provided comprehensive scientific computing, including statistical calculations \cite{bressert2012scipy, scikit}.

\subsection{Missing Value Imputation}

In this study, missing neutron flux measurements for specific assemblies were handled using linear interpolation, selected because flux measurements vary continuously with axial position along the wire, making local interpolation between neighbouring points a reasonable estimation method. This process was applied across all assemblies and historical cycles. Where zero counts occurred at the boundary preceding a missing measurement, interpolation occasionally produced negative values. For such values, the nearest neighbouring measurement was used to ensure the imputed data remained physically realistic.

\subsection{Scanning Direction Correction}

In accordance with the measurement protocol, and to expedite the process, the measurement direction is alternated between top-to-bottom and bottom-to-top for consecutive wires. This should be taken into account when interpreting the locations of measurement points along the wire and the corresponding measurement times.

\subsection{Decay Correction}

Correction of radioactive decay is an essential step in accurately determining the activity yields after measurements \cite{dawood2013correction}. Because the 32 assemblies and 180 data points along each wire are measured sequentially rather than simultaneously, the measured counts correspond to different amounts of elapsed decay time and must be corrected back to a common reference time before the resulting profiles can be meaningfully compared. Copper wires are activated primarily to $\ce{^{64}Cu^{*}}$, which has a half-life of $T_{1/2} = \SI{12.7}{\hour}$ as discussed in Section~\ref{Introduction}. Following the procedure discussed by Kgolobe et al. \cite{Kgolobe2025SAIP}, the decay correction was applied to each assembly measurement and 180 data points along the wires. Specifically, the measured counts, $y$, for a given data point was corrected back to its value at the start of the first measurement using $\tilde{y} = y e^{\lambda \Delta{t}}$, where $\lambda = \text{ln(2)}/T_{1/2}$ and $\Delta{t}$ accounts for both the start time of each assembly and the sequential measurement time ($\SI{1}{\second}$ per point) \cite{dawood2013correction}. Full details of the decay correction process are provided in \cite{Kgolobe2025SAIP}.

\subsection{Polynomial Fitting}

Following decay correction, higher-order polynomial fitting was used to reduce statistical noise and enable consistent relevant flux features. High-order polynomials were fitted to each flux profile following the approach detailed in \cite{Kgolobe2025SAIP}. For each fuel assembly, axial point and operational cycle, a high-order polynomial of degree $p$ was fitted:
\begin{equation}
    P(z) = a_0 + a_{1}z + a_{2}z^{2} + \ldots + a_{p}z^{p},\label{3}
\end{equation}
where $z$ is the axial coordinate with the origin located at the bottom of the wire and the coefficients $a_{0}, a_{1}, \ldots, a_{p}$ are determined by minimising the squared error between the polynomial and the direction- and decay-corrected measurements $\tilde{y}_j$ along the wire. The first derivative of $P(z)$ was used to locate extrema of the axial flux profile: maximum, as well as top and bottom minima, used as key features in this study. Complete details of the fitting procedure and polynomial degree selection are provided in \cite{Kgolobe2025SAIP}.

\subsection{Data Normalisation}

In addition to preprocessing the decay-corrected flux measurements, data normalisation was performed. This is a commonly used technique to rescale numerical features to a standard scale, such as $[0, 1]$ or $[-1, 1]$, ensuring that no single feature dominates the others. In this study, the decay-corrected flux measurements were normalised to standardise the flux magnitudes across assemblies and cycles. A core-average normalisation factor was denoted as $\zeta_{c}$, for each operational cycle $c$. The decay-corrected flux measurements were then normalised using their corresponding core average as:
\begin{equation}
    \bar{y}^{(i)}_{j,c} = \frac{\tilde{y}^{(i)}_{j,c}}{\zeta_{c}},
    \label{6}
\end{equation}
where $\tilde{y}_{i, j, c}$ is the decay corrected flux measurements for each assembly $i$, axial point $j$ and operational cycle $c$. This process ensures that the neutron flux measurements are scaled to a common core scale, which is essential for the ML modelling.

In addition, both the axial and control bank positions were normalised using the Min-Max scaler, a popular data normalisation technique that maps data to a specific range, usually $[0, 1]$. The Min-Max scaler was applied separately to the axial positions and the control bank positions, transforming both to a common range.

\subsection{Neutron flux Feature Extraction}

In addition to the polynomial fitting of neutron flux measurements, three extremum points of the polynomial fit for fuel assemblies were extracted. The extrema were identified as the points where the first derivative of the fitted polynomial equals zero. The extrema extracted are the bottom minimum, maximum and top minimum points corresponding to the intrinsic physical characteristics of the neutron flux profile, as discussed in Section~\ref{Introduction}. Their relevance as meaningful anomaly indicators is explained and validated in \cite{Kgolobe2025SAIP}.

The maximum points were determined using
\begin{equation}
    z_{{\text{m}},c}^{(i)} = \arg\max_{z} P^{(i)}_{c}(z)\label{13},
\end{equation}
while the bottom and top minimum points were determined using
\begin{equation}
    z_{s,c}^{(i)} = \arg\min_{z \lessgtr z_{\text{m},c}^{(i)}} P_c^{(i)}(z), \quad s \in \{\text{b}, \text{t}\},
    \label{eq:minima}
\end{equation}
where ``$<$'' applies for $s = \text{b}$ (bottom minimum) and ``$>$'' applies for $s = \text{t}$ (top minimum). Hence, $z_{\text{m},c}^{(i)}$, $z_{\text{b},c}^{(i)}$, and $z_{\text{t},c}^{(i)}$ represent the maximum, bottom minimum, and top minimum points, respectively, for assembly $i$ and operational cycle $c$.

\subsection{Skewness of Flux Shapes}

In addition to the above-mentioned flux profile feature, the skewness of the flux shapes was investigated. Skewness of the flux shapes serves as an additional feature in the ML inputs to capture their asymmetry. The main objective is to obtain features of the flux profiles that will indicate whether the flux profile is symmetric or asymmetric, to indicate the distorted flux profiles \cite{Demir2022comparison}. The skewness of the flux shapes was calculated over the interval bounded by the bottom and top minimum points $K^{(i)}_{c} = \left\{ j \ \middle| \ z_j \in \left[z_{\text{b},c}^{(i)},\ z_{\text{t},c}^{(i)}\right] \right\}$. A linear reference line $\hat{y}$ was fitted over the neutron flux measurements over this interval:
\begin{equation}
    \hat{y}^{(i)}_{c}(z) = m^{(i)}_{c} z + b^{(i)}_{c} \quad \text{for} \quad z \in K^{(i)}_{c}\label{16},
\end{equation}
where $\hat{y}^{(i)}_{c}(z)$ represents the fitted reference line between the bottom and top minima, and  $m^{(i)}_{c}$ and $b^{(i)}_{c}$ are its slope and intercept, respectively. The fitted reference line was then used to adjust the neutron flux measurements by subtracting $\hat{y}^{(i)}_{c}(z)$ from the corresponding measurements and thereby separate local asymmetries pertinent to the skewness calculation:
\begin{equation}
    \delta y_{j,c}^{(i)} = \tilde{y}_{j,c}^{(i)} - \hat{y}_{c}^{(i)}(z_j),
    \label{17}
\end{equation}
where $\tilde{y}_{j,c}^{(i)}$ are the decay-corrected neutron flux measurements over the interval $K^{(i)}_{c}$ and $\delta y_{j,c}^{(i)}$ are the adjusted neutron flux measurements for assembly $i$, cycle $c$ and point $j$. Furthermore, discrete normalised distributions $p^{(i)}_{j,c}=p^{(i)}_{c}(z_j)$, where $z_j \in K^{(i)}_{c}$, were calculated for the adjusted neutron flux measurements as:
\begin{equation}
    p^{(i)}_{j,c} = \frac{\left|\delta y_{j,c}^{(i)}\right|}{\sum_{j' \in K^{(i)}_{c}} \left|\delta y_{j',c}^{(i)}\right|},
    \label{18}
\end{equation}
where $p^{(i)}_{j,c}$ are the discrete normalised distributions of the neutron flux measurements and $\sum_{j \in K^{(i)}_{c}} p^{(i)}_{j,c} = 1$. The absolute value is applied to ensure $p^{(i)}_{j,c}$ is non-negative and forms a valid discrete distribution; the directional information required for the skewness calculation is preserved separately through the signed deviation in Equation~\eqref{21}. The mean and standard deviation of this distribution were used to calculate the skewness of the flux shapes as:
\begin{equation}
        \gamma^{(i)}_{c} = \frac{\sum_{j \in K^{(i)}_{c}} \left(\delta y_{j,c}^{(i)} - \mu^{(i)}_{c}\right)^3 p^{(i)}_{j,c}}{\left(\sigma^{(i)}_{c}\right)^3}\label{21},
\end{equation}
where $\gamma^{(i)}_{c}$ represents the skewness of the flux shapes for assembly $i$ and cycle $c$. The mean of the axial flux measurements are denoted as $\mu^{(i)}_{c}$ and standard deviation $\sigma^{(i)}_{c}$ for the operational cycles $c$.

\section{DBSCAN For Anomaly Detection}\label{DBSCANForAnomalyDetection}

\subsection{Feature Adjustment for Control Rod Effects}

The neutron flux measurements in the SAFARI-1 research reactor are affected by several factors: these include the current reactor state (including the control rod positions) and its history since fuelling, improper insertion of the copper wire during measurement and statistical noise. In this section, we focus on identifying anomalies caused by improper copper wire insertion, while statistically adjusting the measured flux profiles to eliminate the effects of the other two factors: control rod movement and statistical noise. Statistical noise was handled by the smoothing effect of applying high-order polynomials as discussed in Section~\ref{DataPreprocessing&Feature Extraction}.
Furthermore, to remove the effect of the control rods, the flux features were adjusted to an average reference line, following the established linear relationship between the flux features and control bank positions \cite{Kgolobe2025SAIP}. Each extremum feature $\alpha \in \{\text{m}, \text{b}, \text{t}\}$ (the maximum, bottom minimum, and top minimum points, respectively) defined in Equations~\eqref{13}-\eqref{eq:minima} was adjusted by removing the deviation from this reference line. This adjustment isolates variations caused by improper copper insertion from those caused by the control rod movement:
\begin{equation}
    z_{\text{adj},\alpha}^{(i,c)} = z_{\alpha}^{(i,c)} - \left(\hat{z}_{\alpha}^{(i,c)} - \bar{z}_{\alpha}^{(i)}\right),
    \label{eq:adjustment}
\end{equation}
where $z_{\text{adj},\alpha}^{(i,c)}$ are the flux features that only have improper copper wire insertion for feature $\alpha$ for fuel assembly $i$ and operational cycle $c$. $\hat{z}_{\alpha}^{(i,c)}$ is the fitted value of feature $\alpha$ predicted from the linear relationship with control bank position for cycle $c$; and $\bar{z}_{\alpha}^{(i)}$ is the fixed reference average of feature $\alpha$ for assembly $i$. These two processes account for statistical noise and control rod movement, isolating the variations more closely associated with improperly inserted copper wires.

\subsection{Feature Normalisation}

In line with the recommendation in \cite{Deng2020}, the four features, namely bottom minimum, maximum, top minimum, and skewness of the flux shapes, were normalised using the Min-Max Scaler described in Section~\ref{DataPreprocessing&Feature Extraction}. Furthermore, these four features were combined into a four-dimensional feature matrix for use as input to the ML task.

\subsection{DBSCAN Algorithm}

The DBSCAN clustering algorithm is a density-based clustering technique that groups data points that are `close' to each other into clusters and classifies outliers as noise \cite{hahsler2019dbscan}. The DBSCAN clustering algorithm requires two parameters: the radius of the neighbourhood, $\varepsilon$, and the minimum number of points to form a cluster, \text{MinPts} \cite{hahsler2019dbscan}. In this study, the DSBCAN clustering algorithm was employed to group the neutron flux profiles into those resulting from improper copper wire insertion and those that do not (i.e., anomalous and normal flux profiles). The DBSCAN clustering algorithm was employed using the \texttt{sklearn} library, as mentioned in Section~\ref{DataPreprocessing&Feature Extraction}. The four features mentioned above, the three adjusted extrema features $z_{\text{adj},\alpha}^{(i,c)}$, $\alpha \in \{\text{m}, \text{b}, \text{t}\}$ (Equation~\eqref{eq:adjustment}), together with the skewness feature $\gamma_c^{(i)}$ (Equation~\eqref{21}) were assembled to four-dimensional input feature vectors, $\bm{v}^{(i)}_{c}=(z_{\text{adj},\text{m}}^{(i,c)}, z_{\text{adj},\text{b}}^{(i,c)}, z_{\text{adj},\text{t}}^{(i,c)}, \gamma_c^{(i)})$. These vectors were used as input to perform clustering of the neutron flux profiles. The L$_1$ norm, also known as the Manhattan norm, which calculates the sum of the absolute values of a vector's components \cite{wang2014robust}, was employed to calculate the distance between the input feature vectors:
\begin{equation}
    d^{(i)}_{c,c'} = \sum_{q=1}^{4} \left| v^{(i)}_{c,q} - v^{(i)}_{c',q} \right|,
    \label{eq:l1norm}
\end{equation}
where indices $c$ and $c'$ correspond to flux profiles of two operational cycles being compared.

\subsection{Parameter Selection}

Selecting optimal DBSCAN parameters is crucial, as they directly affect the algorithm's performance \cite{hahsler2019dbscan}. These parameters highly influence each other; hence, changing one parameter requires changing the other. In this study, optimal values of $\varepsilon$ were determined using $k$-distance graphs. For each feature, the distance to its $k$-th nearest neighbour was calculated and sorted in ascending order. The point where the curve reaches its local minimum, referred to as the elbow point, was selected as the optimal $\varepsilon$ value \cite{blogdbscan}. This analysis was performed for each assembly using the Python \texttt{kneed} package, which implements the Kneedle Algorithm to locate the elbow point \cite{blogdbscan}. MinPts values were varied between 10 and 25 to assess their impact on the clustering outcome, with the final values selected based on clustering stability and domain expert validation.

\section{Convolutional Autoencoders for Anomaly Detection}\label{ConvolutionalAutoencodersForAnomalyDetection}

\subsection{Data Preparation and Splitting}

The dataset for CAE development consisted of normalised decay-corrected flux measurements that domain experts had classified. The domain experts used visualisations that included polynomial fits along with the minimum points at the bottom and top of all historical cycles, and tracked points that deviated from the others, classifying them as abnormal profiles. This approach ensured that normal and abnormal flux profiles were separated into two different datasets for further modelling. Among the 92 operational cycles considered in this study, 73, 67, 76, and 61 were identified as normal, and 19, 25, 16, and 31 were identified as abnormal for fuel assemblies B3, D5, E6, and H3, respectively. Furthermore, the normal flux profiles were split into three sets: training, validation, and test. After splitting the normal flux profiles, the test set consisted of the last four cycles, since the cycles were ordered chronologically from oldest to most recent before splitting. The validation set consisted of the six cycles preceding the test cycles, isolated after the test cycles were set aside. Finally, the training set comprised the remaining cycles after both validation and test sets were isolated, totalling 63, 57, 66, and 51 cycles for fuel assemblies B3, D5, E6, and H3, respectively. This approach ensured that the CAE models were trained only on normal flux profiles, with distinct cycles designated for training, validation and testing within each assembly's dataset. Applying this process separately to each assembly results in four CAE models. In addition to the normal flux profiles, the associated normalised axial and control-bank positions were included as inputs to the CAE models.

\subsection{CAE Architecture}

In this study, we employed four individual CAE models for assemblies B3, D5, E6, and H3 to detect the anomalies in the neutron flux measurements caused by improperly inserted copper wires. These assemblies were selected based on their distinct locations within the core and unique neutron flux distribution. Assemblies D5 and E6 are located near the core centre and exhibit a relatively smooth flux distribution with a pronounced dependence on control bank position. In contrast, assemblies B3 and H3 are positioned at the periphery, where flux distributions contain significant statistical noise and their axial flux profiles are less sensitive to the control rod movement.

\begin{figure}[htb!]
  \centering
    \includegraphics[width=0.99\textwidth]{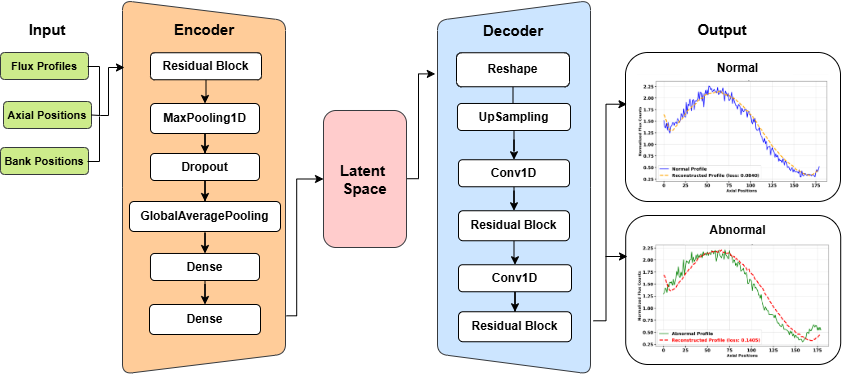}
  \caption {Schematic of the CAE architecture with inputs (left), encoder (orange), latent space (middle), decoder (blue) and the two outputs normal or abnormal flux profiles (right).}
  \label{fig:Architecture_model}
\end{figure}

The proposed models were developed using the \texttt{Keras} API, part of the \texttt{TensorFlow} platform, which provides a straightforward and effective framework for solving ML problems \cite{Keras}. For each assembly considered, the CAE encoder compresses the input flux profile into a latent-space representation capturing its underlying characteristics. At the same time, the decoder reconstructs the flux profile from this representation. This encoder-decoder structure forms the basis for identifying anomalies caused by improperly inserted copper wires in the flux profiles through reconstruction errors, as described in the following section. Figure~\ref{fig:Architecture_model} illustrates the proposed CAE architecture with the input and output data alongside the encoder and decoder, and finally, on the right are the visualised outcomes of the reconstructed neutron flux measurements.

The input sequence for assembly $i$ and cycle $c$, comprising the normalised decay-corrected measurements, axial positions, and control bank positions across all axial points along the wire, is encoded into a latent vector as follows:
\begin{equation}
L^{(i)}_{c} = f_{\text{enc}}\left( X^{(i)}_{c} \right) \label{58},
\end{equation}
where $X^{(i)}_{c}$ is the input sequence for assembly $i$ and cycle $c$, $f_{\text{enc}}$ is the encoder mapping, and $L^{(i)}_{c}$ is the resulting latent vector for assembly $i$ and cycle $c$.

The decoder reconstructs the neutron flux profile from this latent representation, mirroring the encoder structure in reverse:
\begin{equation}
r^{(i)}_{c} = f_{\text{dec}}\left( L^{(i)}_{c} \right) \label{59},
\end{equation}
where $f_{\text{dec}}$ is the decoder mapping, and $r^{(i)}_{c}$ is the resulting reconstructed output for assembly $i$ and cycle $c$. $r^{(i)}_{c}$ is a sequence of 180 axial points, at each of which the decoder reconstructs three values, corresponding to the normalised flux measurement, axial position, and control bank position supplied as input at that point. Since only the flux measurement is relevant for detecting anomalies in this study, the reconstructed flux measurement channel at axial point $j$ is extracted from $r^{(i)}_{c}$ and denoted $r^{(i)}_{j,c}$.



\subsection{Loss Function}

As part of training the CAE models, a customised weighted mean absolute error (MAE) loss function was applied to emphasise accurate reconstruction of the physically significant minima regions. This approach computes the difference between the input and reconstructed flux profiles. During training, the proposed loss function assigns greater weights to essential regions such as regions near the bottom and top ends of the neutron flux profiles, where anomalies caused by improperly inserted copper wires typically occur, as discussed in Section~\ref{Problem_Definition}. This approach ensures that the models correctly learn the characteristics of the flux profiles in these regions. By learning these characteristics accurately, the model is better able to flag deviations from them as anomalies. The loss function was employed as follows:
\begin{equation}
    \mathcal{L}_{\text{MAE}}^{(c)} = \frac{1}{180} \sum_{j=1}^{180} w_j \left| \tilde{y}^{(i)}_{j,c} - r^{(i)}_{j,c} \right|\label{60},
\end{equation}
where $w_j$ represents the weights assigned to the axial position $j$ enumerated from bottom to top). The weight decreases linearly from 3 to 1 over the first $\SI{10}{\percent}$ of the axial points, remains fixed at 1 over the central $\SI{70}{\percent}$ of the points, and increases linearly from 1 to 3 over the final $\SI{20}{\percent}$ of the points. This approach ensures that the model prioritises accurate reconstruction near the ends of the profile, where the flux distribution arising from the surrounding materials, such as water around the core, as discussed in Section~\ref{Introduction}.

\subsection{Hyperparameter Optimisation}
The performance of ANN models depends critically on the selection of optimal model parameters. Systematic model parameter selection approaches have been developed to identify optimal configurations without manual selection. Techniques such as  Grid Search, Random Search (RS) and Bayesian optimisation have been widely used to optimise model hyperparameters during training \cite{Andonie2019hyperparameter}.
In this study, hyperparameter optimisation was performed using 5-fold cross-validation to identify the model configuration yielding the best performance, and $F_1$ scores were evaluated across folds. Given the size of the hyperparameter search space defined in Table~\ref{tab:hyperparameters}, an RS approach was employed, in which each hyperparameter value was independently sampled from its predefined parameter range in Table~\ref{tab:hyperparameters}. For each sampled configuration, the average $F_1$ score across the five folds was computed, and the configuration yielding the highest mean score was selected for final model training.

\begin{table}[htb]
\centering
\small
\renewcommand{\arraystretch}{1.1}
\caption{Hyperparameter values considered during random Search Cross-Validation}
\label{tab:hyperparameters}
\begin{tabular}{lcc}
    \toprule
    Hyperparameters              &      Tested Values      &  \\ \midrule
    First residual block filter  &      [32, 64, 128]      &  \\
    Second residual block filter &     [64, 128, 256]      &  \\
    Third residual block filter  &     [128, 256, 512]     &  \\
    Dense units                  &     [64, 128, 256]      &  \\
    Latent size                  &     [360, 450, 720]     &  \\
    Dropout rate                 &     [0.1, 0.2, 0.3]     &  \\
    Weight decay                 &  [0.0001, 0.001, 0.01]  &  \\
    Learning rate                & [0.0001, 0.0005, 0.001] &  \\
    Batch size                   &      [16, 32, 64]       &  \\
    Epochs                       &     [200, 300, 400]     &  \\
    Patience                     &      [15, 25, 35]       &  \\ \bottomrule
\end{tabular}
\end{table}

\subsection{Model Training}
AdamW (Adam with decoupled weight decay) was employed during both hyperparameter tuning and model training to enhance performance by decoupling weight decay from the gradient update step \cite{loshchilov2017fixing}. Early stopping was employed by monitoring validation loss and halting training when no improvement was observed for the specified patience period. This approach ensures that unnecessary training does not occur and mitigates overfitting by stopping training when validation performance stops improving. Additionally, the learning rate reduction parameter was applied when the validation loss stopped improving for a specified number of epochs.

\subsection{Anomaly Detection and Threshold Selection}
After training the models with optimal parameters, reconstruction errors were computed for all cycles (both normal and abnormal profiles). The percentile method was used to select a threshold to separate normal from abnormal. The percentile method is a simple yet powerful technique that sets thresholds based on the distribution of data, where a specific threshold is defined as the point below which a given percentage of the data falls \cite{komadina2024comparing}. Using the percentile method, candidate thresholds were evaluated between the 90th and 99th percentiles of the normal reconstruction error distribution. Furthermore, for each threshold a confusion matrix was computed, along with the precision, recall and $F_1$ scores as indicated by Table~\ref{tab:metrics}.

\begin{table}[htb]
\centering
\small
\renewcommand{\arraystretch}{1.1}
\caption{Evaluation metrics used for anomaly detection performance assessment}
\label{tab:metrics}
\begin{DIFnomarkup}
\begin{tabular}{p{2.25cm}p{2.75cm}p{8.5cm}}
    \toprule
    Evaluation Metric & Formula                                  & Definition                                                                                                        \\ \midrule
    Precision ($P$)   & $P = \dfrac{\text{TP}}{\text{TP} + \text{FP}}$                & Proportion of correctly identified anomalies among all instances classified as anomalous~\cite{hossin2015review}. \\ \midrule
    Recall ($R$)      & $R = \dfrac{\text{TP}}{\text{TP} + \text{FN}}$                & Proportion of actual anomalies that were correctly detected by the model~\cite{hossin2015review}.                 \\ \midrule
    $F_1$ Score       & $F_1 = 2 \cdot \dfrac{P \cdot R}{P + R}$ & Harmonic mean of precision and recall, providing a balanced measure of model performance~\cite{hossin2015review}. \\ \bottomrule
    where, &  & \tabularnewline
    \multicolumn{3}{l}{TP (True Positive): the model predicts an anomaly case, and this is correct.}\tabularnewline
    \multicolumn{3}{l}{FP (False Positive): the model predicts an anomaly case, but this is wrong (false alarm).}\tabularnewline
    \multicolumn{3}{l}{FN (False Negative): the model predicts a regular case, but this is wrong (missed detection).}\tabularnewline
    \multicolumn{3}{l}{TN (True Negative): the model predicts a regular case, and this is correct.}\tabularnewline
\end{tabular}
\end{DIFnomarkup}
\end{table}

The final threshold $T^*$ was selected as the value that maximised $F_1$ scores while maintaining a balance between precision and recall for anomaly detection. The final threshold was computed as follows:
\begin{equation}
    T^{*} = \arg \max_{T}{F_1}(T)\label{22}.
\end{equation}
Equation~\eqref{22} ensures that the selected threshold that separates between normal and abnormal is the one that results in the best performance as guided by the $F_1$ score.

\section{Results and Discussion} \label{Results&Discussion}

This section presents the results of the clustering and CAE approaches for detecting anomalies caused by improperly inserted copper wires. The clustering-based DBSCAN algorithm was applied first, using four features to identify outliers. Fuel assemblies D5 and H3 were selected to evaluate DBSCAN's performance based on their locations within the core, D5 near the centre and H3 at the periphery, as in Figure~\ref{fig:SAFARI-1-top-view}. The more advanced CAE-based approach was then applied, with hyperparameters tuned using cross-validation random search to identify the best-performing model configurations. Four assemblies were selected for this evaluation: B3 (periphery, south of the core), D5 (near the centre), E6 (centre, between control rods), and H3 (periphery, adjacent to the poolside facility), enabling the feasibility of the CAE models to be assessed across a range of core locations.

\subsection{DBSCAN Results}

Prior to clustering, the effects of statistical noise and control rod movement were addressed. Statistical noise was handled by applying a high-order polynomial fit, and the effect of control rod movement was accounted for by adjusting the extracted features to a calculated average reference line. Optimal DBSCAN parameters $\varepsilon$ and MinPts were determined using $k$-distance curves for core positions D5 and H3. The determined $\varepsilon$ values ranged between 0.2 and 0.4, with \text{MinPts} varied between 10 and 25, as detailed in Table~\ref{tab:paramter_epsilon}.

\begin{figure}[htb]
    \centering
    \begin{minipage}[b]{0.495\textwidth}
        \includegraphics[width=\linewidth]{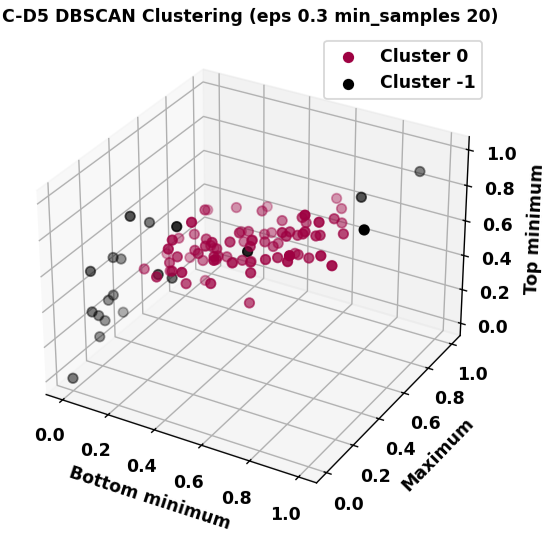}
    \end{minipage}%
    \hfill
    \begin{minipage}[b]{0.495\textwidth}
        \includegraphics[width=\linewidth]{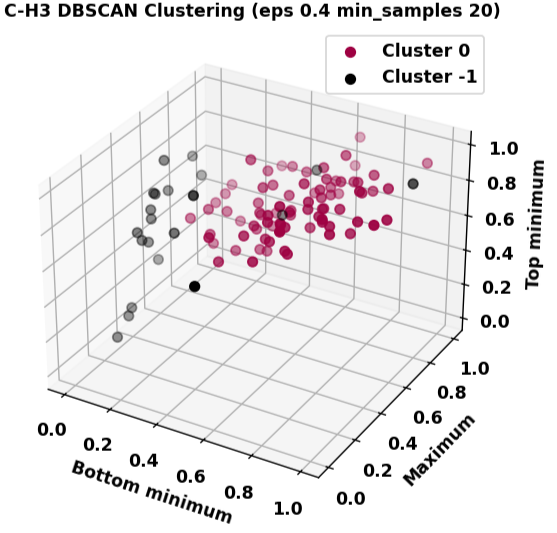}
    \end{minipage}%
    \caption{Clustered features using the DBSCAN algorithm for assembly D5 (left) and H3 (right).}
    \label{fig:DBSCAN}
\end{figure}

\begin{DIFnomarkup}
\begin{table}[htb]
\centering
\small
\renewcommand{\arraystretch}{1.1}
\caption{DBSCAN parameter values for different core positions}
\label{tab:paramter_epsilon}
\begin{minipage}{0.45\textwidth}
\centering
\begin{tabular}{lcc}
    \hline
    Core Position & $\varepsilon$ & MinPts \\ \midrule
    B3             &      0.4      &      10      \\
    B4             &      0.4      &      15      \\
    B5             &      0.3      &      25      \\
    B7             &      0.3      &      15      \\
    C4             &      0.3      &      10      \\
    C6             &      0.3      &      20      \\
    C8             &      0.4      &      20      \\
    D3             &      0.4      &      15      \\
    D4             &      0.3      &      15      \\
    D5             &      0.3      &      20      \\
    D7             &      0.4      &      20      \\
    E4             &      0.4      &      20      \\
    E6             &      0.2      &      20      \\ \bottomrule
\end{tabular}
\end{minipage}
\begin{minipage}{0.45\textwidth}
\centering
\begin{tabular}{lcc}
    \toprule
    Core Position & $\varepsilon$ & MinPts \\ \midrule
    E8             &      0.3      &      20      \\
    F3             &      0.3      &      20      \\
    F4             &      0.3      &      20      \\
    F5             &      0.2      &      20      \\
    F7             &      0.3      &      20      \\
    G4             &      0.4      &      20      \\
    G6             &      0.3      &      20      \\
    G8             &      0.4      &      20      \\
    H3             &      0.4      &      20      \\
    H4             &      0.4      &      20      \\
    H5             &      0.3      &      20      \\
    H6             &      0.3      &      15      \\
    H7             &      0.3      &      15      \\ \bottomrule
\end{tabular}
\end{minipage}
\end{table}
\end{DIFnomarkup}
Clustering results, visualised in three dimensions, as illustrated by Figure~\ref{fig:DBSCAN}, show pink points representing flux features that correspond to flux profiles with correctly inserted copper wires and black points (Cluster -1) as outliers, indicating improper insertion. Most of these outliers were associated with improperly inserted copper wires; however, three anomalous flux profiles were also identified for which the underlying causes could not be determined, even though their profiles appeared normal. For assemblies near the core centre (D5), smaller parameter values effectively separated normal and abnormal flux profiles. In contrast, peripheral assemblies H3 required larger parameter values to accommodate the higher statistical noise in these locations. The neutron flux profiles corresponding to the outlier features (black points) were visualised as shown in Figure~\ref{fig:DBSCAN}. Domain experts confirmed that these neutron flux profiles were outliers indicative of anomalies. These profiles were characterised by a secondary peak near the top minimum and a missing peak before the bottom minimum. This analysis highlights DBSCAN's performance in detecting anomalies caused by copper wire insertion, supported by measured data and expert validation.

While the clustering approach demonstrated promising results, its effectiveness was constrained by the input features. The algorithm, which uses features such as polynomial extrema and skewness, could also identify deviations which are rather caused by normal reactor operations. For example, deviations in the maximum key feature points can indicate control rod movements rather than improper wire insertion, which were occasionally flagged as abnormal even though the flux profile is normal. This limitation indicates a fundamental challenge: unsupervised learning methods require careful feature engineering to distinguish operational variations from anomalies \cite{sun2024unsupervised}. Motivated by this limitation, a more expressive, learned approach, namely the CAE, was subsequently applied to the same problem to capture the physical structure of the flux profiles more comprehensively than sets of extracted features could. Even so, when combined with clear preprocessing and feature extraction, DBSCAN demonstrated that unsupervised learning is a feasible approach for anomaly detection.

\subsection{CAE Implementation}
\subsubsection{Hyperparameter Tuning}
The second unsupervised learning technique, applied to detect anomalies caused by improperly inserted copper wires, was CAE. The optimal parameters of the CAE were determined using a random search, which identified hyperparameter combinations yielding the highest mean $F_1$ scores across validation folds ($k = 5$). The tuned architectures and training parameters varied significantly across assemblies, highlighting differences in their flux distributions. For instance, D5 and H3 required deeper networks with larger filter sizes and larger latent spaces compared to E6 and B3. Furthermore, model convergence and the training and validation loss dynamics across assemblies differ. Table~\ref{tab:resulting_parameters} summarises the optimised hyperparameters for each assembly.
\begin{table}[htb]
\centering
\small
\renewcommand{\arraystretch}{1.1}
\caption{Optimised hyperparameters for B3, D5, E6 and H3 assemblies.}
\label{tab:resulting_parameters}
\begin{tabular}{lcccc}
    \toprule
    Hyperparameters            &   D5   &  E6   &   B3   &   H3   \\ \midrule
    First residual block filer &   64   &  128  &   32   &  128   \\
    Second residual filter     &  256   &  64   &  128   &   64   \\
    Third residual filter      &  256   &  128  &  128   &  512   \\
    Encoder dense layer        &  256   &  128  &   64   &  256   \\
    Latent size                &  450   &  360  &  180   &  540   \\
    Dropout rate               &  0.3   &  0.1  &  0.1   &  0.1   \\
    Weight decay               & 0.0001 & 0.01  &  0.01  &  0.01  \\
    Learning rate              & 0.0001 & 0.001 & 0.0001 & 0.0005 \\
    Batch size                 &   16   &  32   &   16   &   64   \\
    Epochs                     &  400   &  300  &  200   &  400   \\
    Patience                   &   35   &  15   &   35   &   35   \\ \bottomrule
\end{tabular}
\end{table}

The cross-validation results, presented in Table~\ref{tab:folds}, demonstrated varying levels of stability in mean $F_1$ scores across folds and assemblies. Assemblies B3, D5 and E6 exhibited relatively stable $F_1$ scores, with B3 showing the strongest consistency as presented by Table~\ref{tab:folds}. These narrow ranges of $F_1$ scores across folds suggest that the models B3, D5, and E6 generalise well across different training sets. In contrast, assembly H3 indicated significant fluctuations in $F_1$ scores across folds, particularly in fold 2, where the score dropped to 0.356 as indicated by Table~\ref{tab:folds}. This variability reflects the elevated statistical noise associated with the H3 core position, resulting in a less clear separation between normal and abnormal cycles. These results underscore the challenges of modelling peripheral assemblies, where statistical noise is higher than in interior assemblies.

\begin{table}[htb]
\centering
\small
\renewcommand{\arraystretch}{1.1}
\caption{Cross-validation $F_1$ scores ($k=5$) for each assembly.}
\label{tab:folds}
\begin{tabular}{lccccc}
    \toprule
    Assembly & Fold 1 & Fold 2 & Fold 3 & Fold 4 & Fold 5 \\ \midrule
    B3       & 0.938  & 0.938  & 0.938  & 0.903  & 0.786  \\
    D5       & 0.850  & 0.850  & 0.878  & 0.878  & 0.905  \\
    E6       & 0.696  & 0.815  & 0.846  & 0.800  & 0.769  \\
    H3       & 0.722  & 0.357  & 0.757  & 0.686  & 0.686  \\ \bottomrule
\end{tabular}
\end{table}
\subsubsection{Central Assemblies D5 \& E6}

After hyperparameter tuning and training the CAEs with the best parameters, the models were evaluated. The performances of the CAE models were evaluated using multiple metrics, including training error, reconstruction error distribution, anomaly detection accuracy, and classification performance. The training history for the central assemblies D5 and E6 demonstrated model convergence, with reduced training and validation losses that stabilise over time. This indicates successful learning of neutron flux profiles and reliable model stability.

\begin{table}
\centering
\small
\renewcommand{\arraystretch}{1.1}
\caption{Reconstruction errors and anomaly detection metrics for D5 and E6 assemblies.}
\label{tab:autoencoder_values_cen}
\begin{tabular}{lccccccccc}
    \toprule
    Assembly & \multicolumn{4}{c}{Mean Absolute Error (MAE)} & Threshold & Precision & Recall & $F_{1}$ &  \tabularnewline \cmidrule{2-5}
             & Train & Validation & Test  &     Anomaly      &           &           &        &         &  \tabularnewline \midrule
    D5       & 0.069 &   0.075    & 0.095 &      0.121       &   0.109   &   0.889   & 0.727  &  0.800  &  \\
    E6       & 0.058 &   0.064    & 0.061 &      0.130       &   0.087   &   0.889   & 0.571  &  0.696  &  \\ \bottomrule
\end{tabular}%
\end{table}

\begin{figure}[htb]
    \centering
    \begin{minipage}[b]{0.45\textwidth}
        \includegraphics[width=\linewidth]{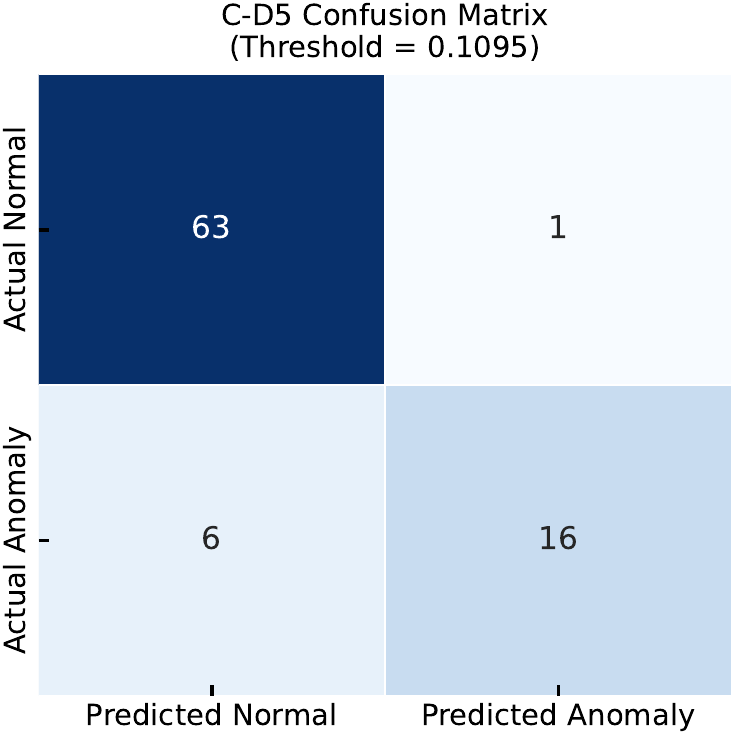}
    \end{minipage}%
    \hfill
    \begin{minipage}[b]{0.45\textwidth}
        \includegraphics[width=\linewidth]{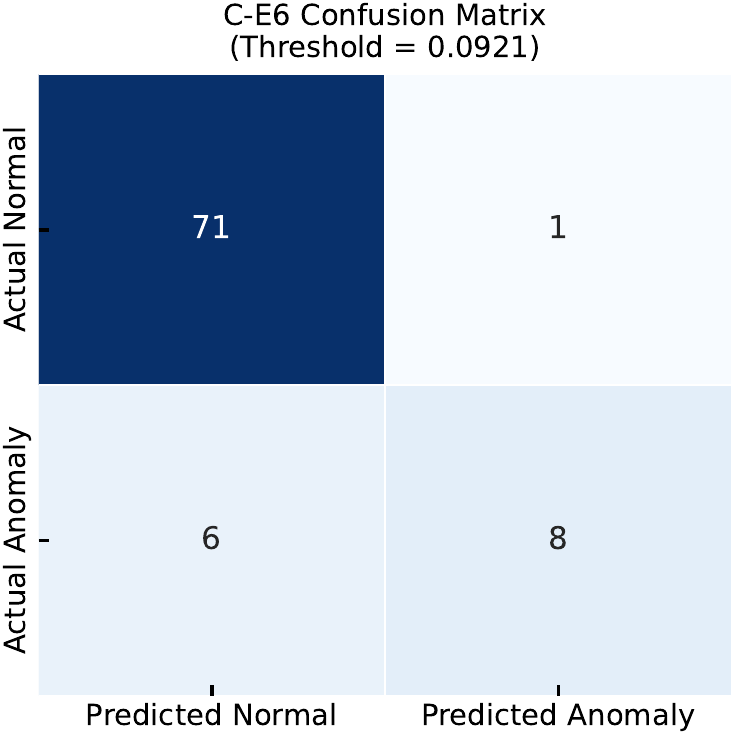}
    \end{minipage}%
    \caption{Confusion matrix plot for core position D5 (left) and E6 (right).}
    \label{fig:confusion_center}
\end{figure}

The anomaly detection metrics for D5 and E6 are summarised in Table~\ref{tab:autoencoder_values_cen}. Assembly D5 achieved an $F_1$ score of 0.800, compared to 0.696 for E6, indicating the ability to effectively distinguish normal from abnormal profiles. The confusion matrices in Figure~\ref{fig:confusion_center} further confirm this finding, with diagonal entries (true positives and negatives) for both assemblies, indicating an accurate classification of neutron flux profiles in most cases. The reconstruction error distributions for D5 and E6 demonstrated that the CAE models effectively learned normal flux profiles, as data points clustered within a narrow range of low reconstruction errors.

\begin{figure}[htb]
\centering
\begin{minipage}[t]{0.49\textwidth}
    \includegraphics[width=\linewidth]{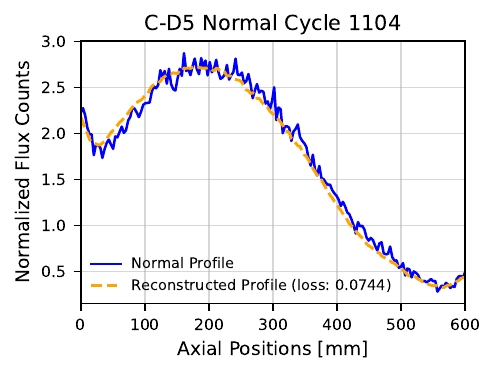}\par\vspace{0.5em}\includegraphics[width=\linewidth]{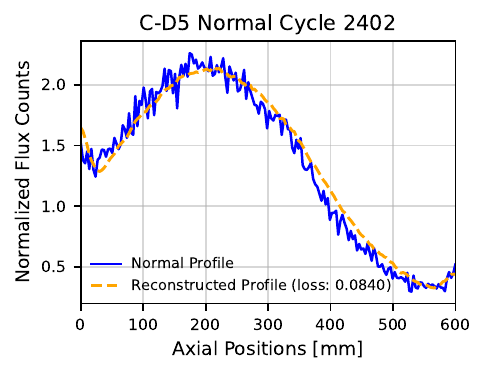}
\end{minipage}
\begin{minipage}[t]{0.49\textwidth}
    \includegraphics[width=\linewidth]{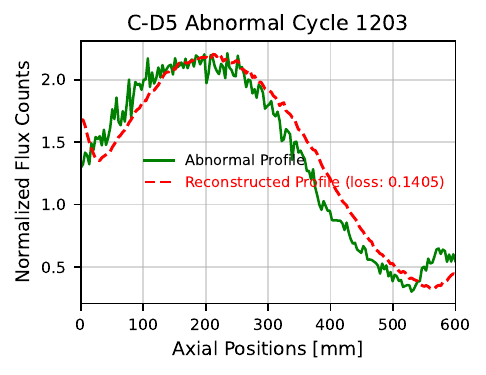}\par\vspace{0.5em}\includegraphics[width=\linewidth]{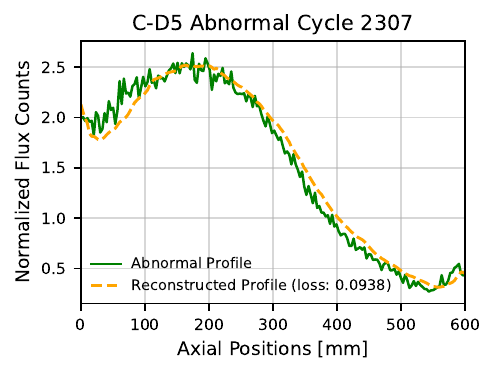}
\end{minipage}
\caption{CAE reconstruction plots for core position D5 for normal (left) and abnormal (right) profiles.}
\label{fig:recons_center_D5}
\end{figure}

\begin{figure}[htb]
\centering
\begin{minipage}[t]{0.49\textwidth}
    \includegraphics[width=\linewidth]{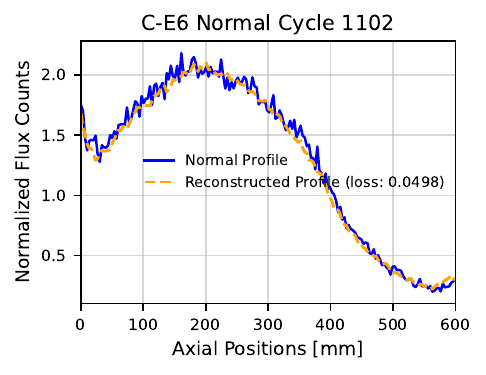}\par\vspace{0.5em}\includegraphics[width=\linewidth]{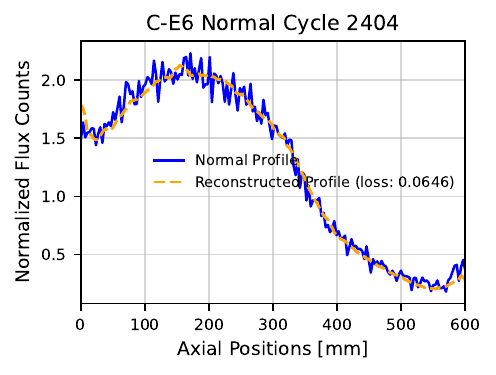}

\end{minipage}
\begin{minipage}[t]{0.49\textwidth}
    \includegraphics[width=\linewidth]{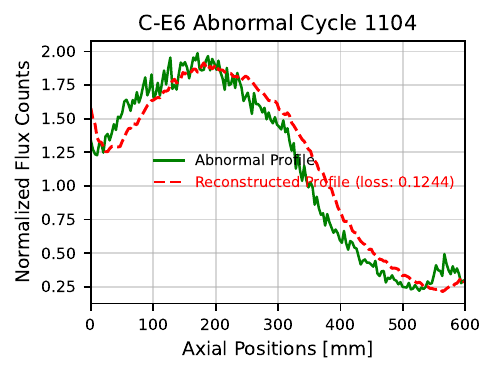}\par\vspace{0.5em}\includegraphics[width=\linewidth]{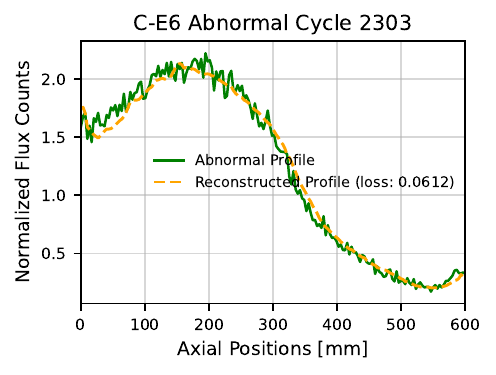}
\end{minipage}
\caption{CAE reconstruction plots for core position E6 for normal (left) and abnormal (right) profiles.}
\label{fig:recons_center_E6}
\end{figure}

Furthermore, a comparison of ground truth and reconstructed measurements is presented in Figures~\ref{fig:recons_center_D5} and \ref{fig:recons_center_E6}. While both assemblies demonstrated excellent reconstruction capabilities and classification of abnormal measurements, discrepancies were noted in specific cycles. For instance, cycle 2307 for assembly D5 was classified as abnormal by domain experts but as normal by the model, and cycle 2303 for E6 was classified as abnormal by experts but as normal by the model. These cases highlight difficult instances in which the anomaly classification was unclear even to domain experts.

\subsubsection{Peripheral Assemblies B3 \& H3}
The performance of CAE models for the peripheral assemblies B3 and H3, located at the periphery of the reactor core, was evaluated. These assemblies exhibit a noisier flux distribution than central assemblies because their lower local flux magnitude results in fewer measured counts. Since relative statistical error is inversely proportional to the square root of the number of counts, this lower count level produces greater statistical noise at peripheral positions. The training history for both B3 and H3 showed fluctuating model convergence, which stabilised during the learning process. The reconstruction and anomaly detection metrics for B3 and H3 are presented in Table~\ref{tab:autoencoder_values_per}. The B3 model achieved a higher score than H3, which obtained the lowest score among all evaluated assemblies. Anomaly detection metrics indicated higher precision but lower recall for B3, while H3 showed comparable precision but significantly lower recall.
\begin{table}[htb]
\centering
\renewcommand{\arraystretch}{1.1}
\small
\caption{Reconstruction errors and anomaly detection metrics for B3 and H3 assemblies.}
\label{tab:autoencoder_values_per}
\begin{tabular}{lcccccccc}
    \toprule
    Assembly  & \multicolumn{4}{c}{Mean Absolute Error (MAE)} & Threshold  & Precision  & Recall  & $F_{1}$ \tabularnewline
    \cmidrule{2-5}
    & Train  & Validation  & Test  & Anomaly  &  &  &  & \tabularnewline
    \midrule
    B3  & 0.054  & 0.054  & 0.069  & 0.085  & 0.0682  & 0.692  & 0.562  & 0.621 \tabularnewline
    H3  & 0.046  & 0.0426  & 0.042  & 0.064  & 0.066  & 0.673  & 0.500  & 0.579 \tabularnewline
    \bottomrule
\end{tabular}%
\end{table}
\begin{figure}[htb]
    \centering
    \begin{minipage}[b]{0.45\textwidth}
        \includegraphics[width=\linewidth]{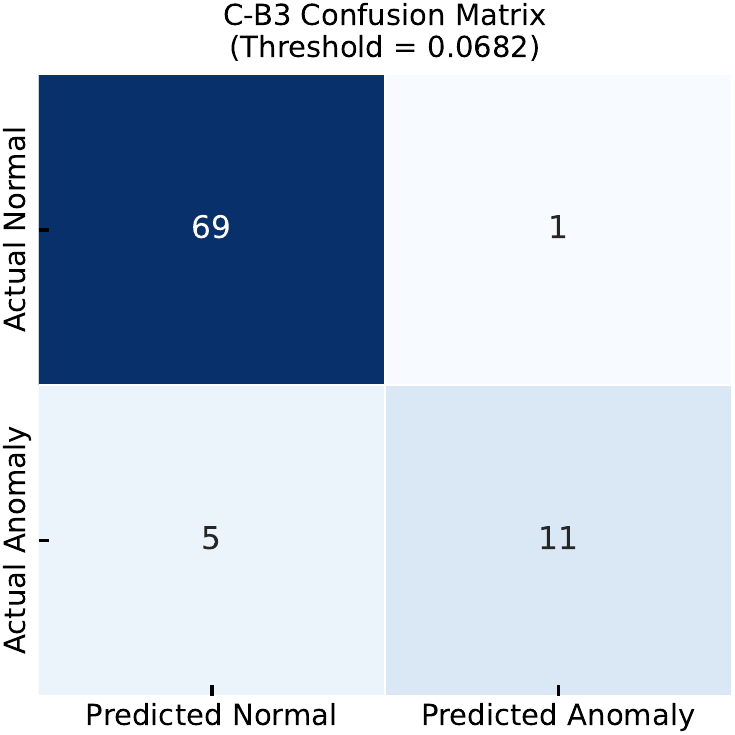}
    \end{minipage}%
    \hfill
    \begin{minipage}[b]{0.45\textwidth}
        \includegraphics[width=\linewidth]{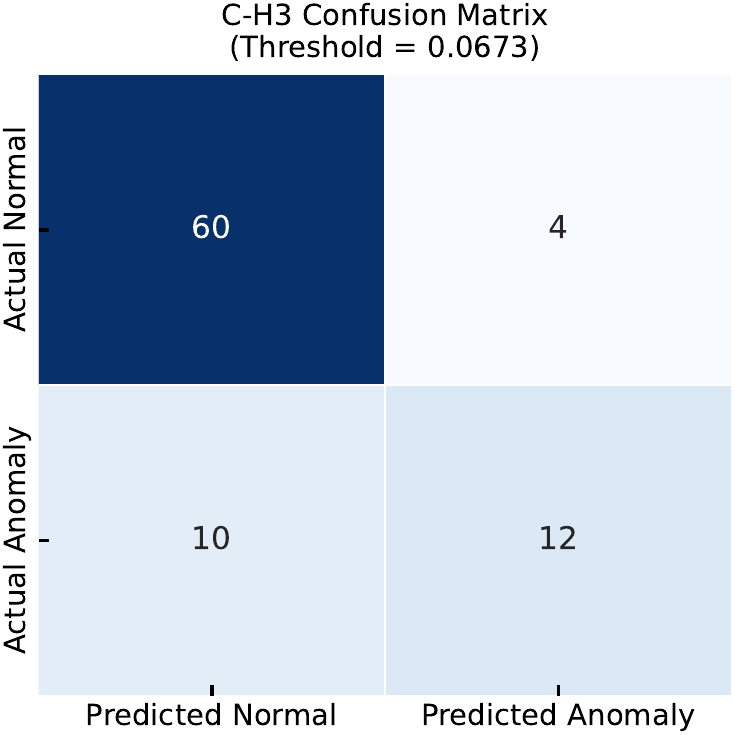}
    \end{minipage}%
    \caption{Confusion matrix plot for core position B3 (left) and H3 (right)}
    \label{fig:confusion_peri}
\end{figure}%

The confusion matrices presented in Figure~\ref{fig:confusion_peri} provide a visual summary of the model's performance. For B3, true positives and true negatives were prominent, indicating robust classification accuracy. In contrast, H3 showed a higher false-negative rate, highlighting challenges in detecting anomalies in noisy peripheral flux profiles.

\begin{figure}[htb]
\centering
\begin{minipage}[t]{0.49\textwidth}
        \includegraphics[width=\linewidth]{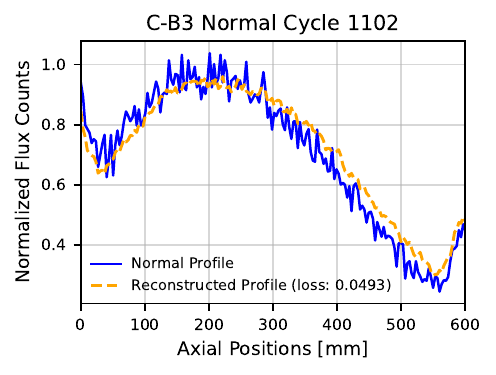}\par\vspace{0.5em}\includegraphics[width=\linewidth]{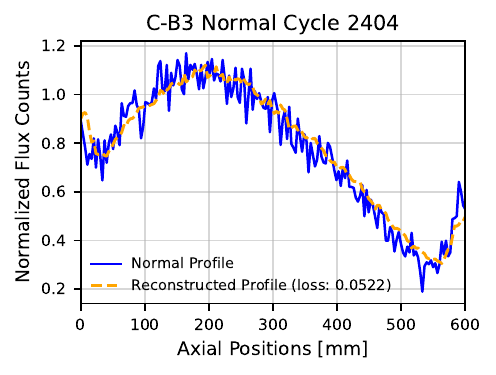}
\end{minipage}
\begin{minipage}[t]{0.49\textwidth}
    \includegraphics[width=\linewidth]{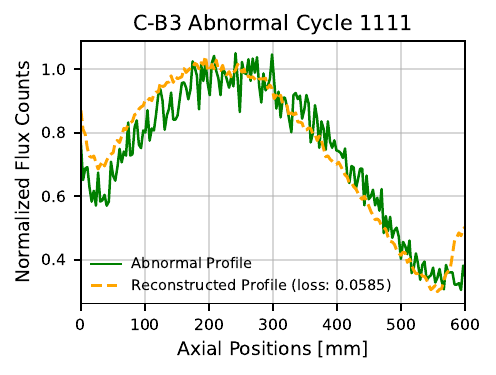}\par\vspace{0.5em}\includegraphics[width=\linewidth]{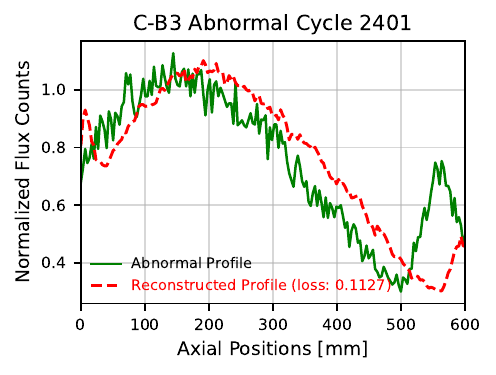}
\end{minipage}
\caption{CAE reconstruction plots for core position B3 for normal (left) and abnormal (right) profiles.}
\label{fig:Recons_B3}
\end{figure}

\begin{figure}[htb]
\centering
\begin{minipage}[t]{0.49\textwidth}
    \includegraphics[width=\linewidth]{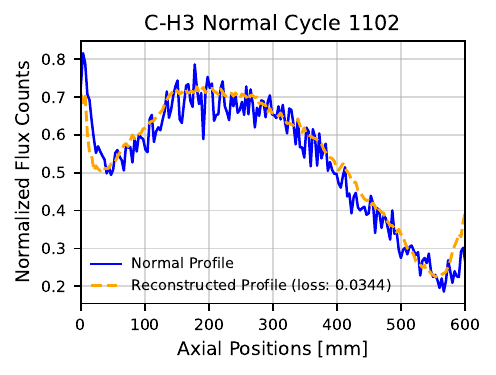}\par\vspace{0.5em}\includegraphics[width=\linewidth]{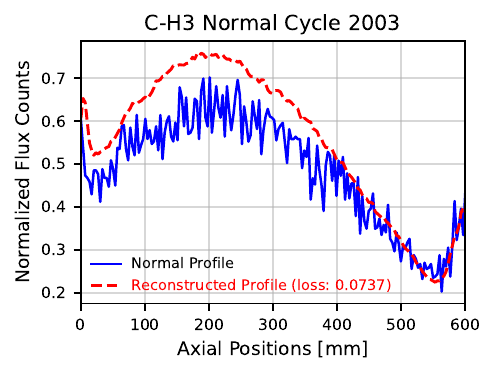}

\end{minipage}
\begin{minipage}[t]{0.49\textwidth}
    \includegraphics[width=\linewidth]{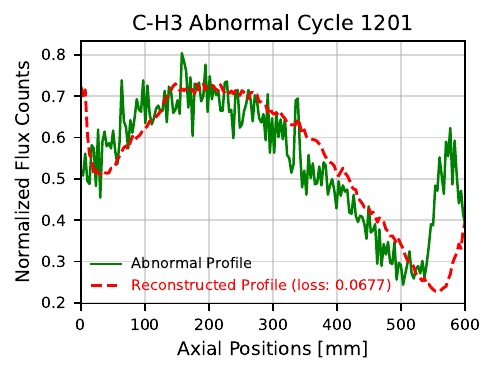}\par\vspace{0.5em}\includegraphics[width=\linewidth]{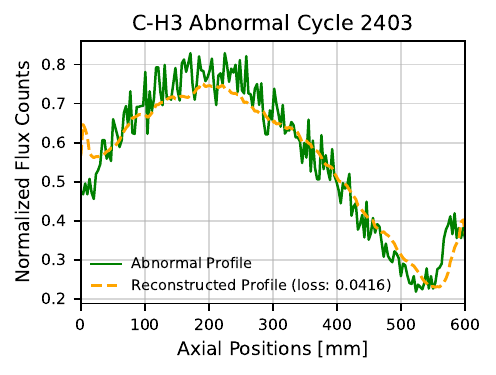}
\end{minipage}
\caption{CAE reconstruction plots for core position H3 for normal (left) and abnormal (right) profiles.}
\label{fig:Recons_H3}
\end{figure}

The reconstruction error distributions for B3 and H3 were wider than those for central assemblies, consistent with the increased statistical noise in peripheral flux measurements. A comparison of reconstruction results in Figures~\ref{fig:Recons_B3} and~\ref{fig:Recons_H3} illustrates that the B3 model accurately reconstructed normal profiles with low errors but misclassified one abnormal cycle (1111) as normal. For H3, the model demonstrated high variability due to increased noise, misclassifying normal cycle (2003) as abnormal and failing to detect one abnormal cycle (2403). These results underscore the impact of statistical noise on model performance for peripheral assemblies.

\subsubsection{Comparison Across Core Positions}
A clear trend emerged when comparing the models' performance across the core, central, and peripheral positions: the tuned CAEs performed better at central assemblies than at peripheral ones. This pattern highlights the dependence of data-driven model accuracy on operational conditions, which vary significantly with assembly location and control rod positions. Central assemblies benefit from higher flux values and clearer signal characteristics, thereby enabling clearer separation between normal and abnormal profiles and enabling more effective learning of the neutron flux profiles.

The findings also indicate that the effectiveness of anomaly detection approaches in nuclear diagnostics requires careful consideration of location-specific characteristics. The architectural differences identified during hyperparameter tuning, with central assemblies requiring moderate complexity and peripheral assemblies demanding deeper architectures, highlight that no single model configuration is optimal across all core positions. Location-based model tuning proved essential for achieving acceptable performance, particularly in the variability in flux distributions across the reactor core.

Considered together, the DBSCAN and CAE results point to a consistent conclusion: both the DBSCAN and CAEs highlighted the potential of data-driven methods for detecting anomalies caused by improperly inserted copper wires, provided they are supported by careful preprocessing, feature extraction and clear separation between normal and abnormal flux profiles. The CAE approach offered advantages over the clustering-based approach by learning hierarchical representations directly from flux profiles, though it remained sensitive to noise in peripheral assemblies. The difference in performance across core positions highlights that future applications of data-driven anomaly detection in nuclear diagnostics should incorporate location-specific considerations into both model development and deployment strategies.

\section{Conclusion and Recommendations}\label{Conclusiona&Recommendations}

This study addressed the detection of anomalies in the neutron flux profiles caused by improperly inserted copper wires in the SAFARI-1 research reactor. By considering both data preprocessing and data-driven algorithms, specifically DBSCAN clustering and CAE-based approach, this work demonstrates that the physical characteristics of measured flux profiles can be leveraged to automate anomaly detection. It also offers a viable complement to the manual inspection methods currently used to support reactor operations. The DBSCAN clustering algorithm demonstrated effective anomaly detection for fuel assemblies near the core centre, while peripheral assemblies required more complex model parameters. Misclassification occurred when deviations were limited to maximum feature variations, indicating reduced sensitivity to isolated feature anomalies. Although clustering successfully identified outliers associated with improperly inserted copper wires, three outliers could not be attributed to improper wire insertion. These anomalies had an unknown source and require further investigation. Tuned CAEs achieved comparable performance in central assemblies but displayed reduced robustness for peripheral regions, indicating sensitivity to data quality and spatial variability within the core.

Limitations of this study arise from the exclusion of control assemblies, whose neutron flux profiles exhibit more complex behaviour, which contrasts with those of fuel assemblies and have different flux shapes. This distribution requires a different approach to feature extraction and complicates the application of both statistical and data-driven methods. The absence of labelled profiles for control assemblies further constrains the CAE approach, as domain experts struggle to classify consistent normal and abnormal cases. Data quality also constrains performance, particularly in peripheral fuel assemblies, where low signal-to-noise ratios in copper-wire measurements reduce the reliability of extracted features and degrade model accuracy.

Future work should prioritise addressing the limitations identified in this study. First, methods must be developed to classify the measurements from control assemblies, which were excluded due to their complex flux distributions and the challenges of feature extraction. Alternatively, obtaining labelled measurements for control rod flux profiles from domain experts would enable a comprehensive evaluation of CAE models on these datasets. Improving the accuracy of labels for normal and abnormal cycles is crucial for reliable model training and interpretation. Domain experts could deploy Monte Carlo N-Particle (MCNP) or SERPENT codes to obtain accurate neutron flux profiles that classify them as normal or abnormal.

Additionally, instances when domain experts are not confident of whether the flux profile is anomalous or normal should be removed or isolated from the datasets to avoid complicating the analysis. Enhancing the quality of neutron flux measurements is another critical area for improvement. The marginal data quality in the peripheral regions, attributed to elevated statistical errors and low signal-to-noise ratios, necessitates improvements in measurement statistics. This can be achieved by increasing count times, increasing count rates, or deploying additional detectors and high-rate data-acquisition systems. Such enhancements would enable more robust data processing and analysis. By addressing these limitations and exploring advanced methodologies, future studies could improve the accuracy of anomaly detection in nuclear reactors. Among these options, the improvement of the activated copper wire measurement system is the most suitable approach and should be investigated. The statistics of the central core region are sufficient, while those of the peripheral region are somewhat marginal. This means a modest increase, by even a factor of two, for the statistics, would be beneficial for these automatic methods.

\section*{CRediT authorship contribution statement}\label{seq:CRediT}

\textbf{Rethabile Kgolobe}: {Conceptualisation, Methodology, Software, Formal analysis, Investigation, Writing -- original draft,
Writing -- review \& editing, Visualisation.} \textbf{Pavel M. Bokov}:  {Conceptualization, Methodology, Formal analysis, Visualisation, Writing -- review \& editing}. \textbf{Simon H. Connell}:  {Conceptualization, Methodology, Formal analysis, Writing -- review \& editing}. \textbf{Charis Harley}: {Conceptualization, Methodology, Formal analysis, Writing -- review \& editing}. \textbf{Bongani G. Maqabuka}:  {Methodology, Formal analysis, Writing -- review \& editing}. \textbf{Lesego E. Moloko}:  {Data Preparation, Writing -- review \& editing}.

\section*{Declaration of competing interest}

The authors declare that they have no known competing financial
interests or personal relationships that could have appeared to
influence the work reported in this paper.

\section*{Data availability}

The authors do not have permission to share data.

\section*{Acknowledgments}\label{seq:Acknowledgments}

The authors gratefully acknowledge the RC Group at the University of Johannesburg (Empire Building) for the computing resources throughout this research. The corresponding author acknowledges the financial support provided by Lesedi Nuclear Services through a Master's bursary, which was not specifically associated with the research reported in this article.


\begin{thebibliography}{10}
    \expandafter\ifx\csname url\endcsname\relax
    \def\url#1{\texttt{#1}}\fi
    \expandafter\ifx\csname urlprefix\endcsname\relax\def\urlprefix{URL }\fi
    \expandafter\ifx\csname href\endcsname\relax
    \def\href#1#2{#2} \def\path#1{#1}\fi

    \bibitem{IAEA_TECDOC_1332_2003}
    {International Atomic Energy Agency},
    \href{https://www-pub.iaea.org/MTCD/Publications/PDF/te_1332_web.pdf}{Safety
        margins of operating reactors: Analysis of uncertainties and implications for
        decision making}, Tech. Rep. IAEA-TECDOC-1332, International Atomic Energy
    Agency, Vienna, Austria (Jan. 2003).
    \newline\urlprefix\url{https://www-pub.iaea.org/MTCD/Publications/PDF/te_1332_web.pdf}

    \bibitem{Lamarsh2018reactortheory}
    J.~R. Lamarsh, A.~J. Baratta, {Nuclear Reactor Theory}, 4th Edition, Pearson
    Education, {Inc.}, {Upper Saddle River, NJ}, 2018.

    \bibitem{Kgolobe2025SAIP}
    {Kgolobe, R. and Bokov, P. M. and Connell, S. H. and Harley, C. and Maqabuka,
        B. G. and Moloko, L. E. and Prinsloo, R. H.},
    \href{http://events2.saip.org.za}{{Detecting Anomalies in Measured Thermal
            Neutron Flux Profiles of SAFARI-1 Research Reactor}}, in: {W. A. Horowitz
        (UCT)} (Ed.), {The Proceedings of SAIP2025, the 69th Annual Conference of the
        South African Institute of Physics}, South African Institute of Physics,
    2025, pp. 336--341.
    \newline\urlprefix\url{http://events2.saip.org.za}

    \bibitem{Chapman1965determination}
    G.~T. Chapman,
    \href{https://catalog.hathitrust.org/Record/007841650}{{Determination of
            Thermal-Neutron Flux Distributions in the Bulk Shielding Reactor II by Copper
            Wire Activation Techniques}}, Report ORNL-3758, Oak Ridge National
    Laboratory, Oak Ridge, TN (1965).
    \newline\urlprefix\url{https://catalog.hathitrust.org/Record/007841650}

    \bibitem{Musa2012determination}
    Y.~Musa, Y.~A. Ahmed, Y.~A. Yamusa, I.~O.~B. Ewa,
    \href{https://www.sciencedirect.com/science/article/pii/S0306454912002605}{{Determination
            of radial and axial neutron flux distribution in irradiation channel of
            NIRR-1 using foil activation technique}}, {Annals of Nuclear Energy} 50
    (2012) 50--55.
    \newblock \href
    {http://dx.doi.org/https://doi.org/10.1016/j.anucene.2012.07.014}
    {\path{doi:https://doi.org/10.1016/j.anucene.2012.07.014}}.
    \newline\urlprefix\url{https://www.sciencedirect.com/science/article/pii/S0306454912002605}

    \bibitem{Moloko2021estimation}
    L.~E. Moloko, P.~M. Bokov, K.~N. Ivanov, Estimation of the axial neutron flux
    profiles in the {SAFARI-1} core using artificial neural networks, in:
    {Proceedings of the International Conference on Mathematics and Computational
        Methods Applied to Nuclear Science and Engineering {(M\&C 2021)}, Virtual
        Meeting}, 2021, pp. 1644--1653.
    \newblock \href {http://dx.doi.org/10.13182/xyz-33887}
    {\path{doi:10.13182/xyz-33887}}.

    \bibitem{Moloko2022quantification}
    L.~E. Moloko, P.~M. Bokov, X.~Wu, K.~N. Ivanov, Quantification of neural
    networks uncertainties with applications to {SAFARI-1} axial neutron flux
    profiles, in: {Proceedings of the International Conference on Physics of
        Reactors ({PHYSOR} 2022), {ANS}, Pittsburgh, Pennsylvania, {USA}}, 2022, pp.
    1398--1407.
    \newblock \href {http://dx.doi.org/10.13182/PHYSOR22-37794}
    {\path{doi:10.13182/PHYSOR22-37794}}.

    \bibitem{articleFiase}
    D.~L.~A. Fiase, K.~O. Attah, W.~O. Amponsah, P.~O. Agyeman, Bridging the gap
    between unsupervised and supervised learning, International Journal of
    Innovative Research and Development 14~(12) (2025) 111--121.
    \newblock \href {http://dx.doi.org/10.24940/ijird/2025/v14/i12/DEC25016}
    {\path{doi:10.24940/ijird/2025/v14/i12/DEC25016}}.

    \bibitem{KOLLIAS2022109293}
    S.~Kollias, M.~Yu, J.~Wingate, A.~Durrant, G.~Leontidis, G.~Alexandridis,
    A.~Stafylopatis, A.~Mylonakis, P.~Vinai, C.~Demaziere,
    \href{https://www.sciencedirect.com/science/article/pii/S0306454922003280}{Machine
        learning for analysis of real nuclear plant data in the frequency domain},
    Annals of Nuclear Energy 177 (2022) 109293.
    \newblock \href
    {http://dx.doi.org/https://doi.org/10.1016/j.anucene.2022.109293}
    {\path{doi:https://doi.org/10.1016/j.anucene.2022.109293}}.
    \newline\urlprefix\url{https://www.sciencedirect.com/science/article/pii/S0306454922003280}

    \bibitem{ioannou2021intelligent}
    G.~Ioannou, G.~Tagaris, Thanos~Alexandridis, A.~Stafylopatis, {Intelligent
        Techniques for Anomaly Detection in Nuclear Reactors}, in: {EPJ Web of
        Conferences}, Vol. 247, {EDP Sciences}, 2021, p. 21011.
    \newblock \href {http://dx.doi.org/10.1051/epjconf/202124721011}
    {\path{doi:10.1051/epjconf/202124721011}}.

    \bibitem{chaudhary2024anomaly}
    A.~Chaudhary, J.~Han, S.~Kim, A.~Kim, S.~Choi, Anomaly detection and analysis
    in nuclear power plants, Electronics 13~(22) (2024) 4428.
    \newblock \href {http://dx.doi.org/10.3390/electronics13224428}
    {\path{doi:10.3390/electronics13224428}}.

    \bibitem{garis1998determination}
    N.~S. Garis, I.~Pázsit, U.~Sandberg, T.~Andersson, {Determination of {PWR}
        control rod position by core physics and neural network methods}, {Nuclear
        Technology} 123~(3) (1998) 278--295.
    \newblock \href {http://dx.doi.org/10.13182/NT98-A2899}
    {\path{doi:10.13182/NT98-A2899}}.

    \bibitem{Huang01081994}
    P.~H. Huang, J.~T. Yang,
    \href{https://doi.org/10.13182/NT94-A34984}{{Qualification of the Taiwan
            Power Company’s Pressurized Water Reactor Physics Methods Using
            CASMO-3/SIMULATE-3}}, {Nuclear Technology} 107~(2) (1994) 138--154.
    \newblock \href {http://dx.doi.org/10.13182/NT94-A34984}
    {\path{doi:10.13182/NT94-A34984}}.
    \newline\urlprefix\url{https://doi.org/10.13182/NT94-A34984}

    \bibitem{terman2018determination}
    M.~S. Terman, N.~M. Kojouri, H.~Khalafi, {Determination of control rod
        positions during fuel life-cycle using fixed in-core Self-Powered Neutron
        Detectors of Tehran Research Reactor}, {Nuclear Engineering and Design} 331
    (2018) 68--82.
    \newblock \href {http://dx.doi.org/10.1016/j.nucengdes.2018.02.026}
    {\path{doi:10.1016/j.nucengdes.2018.02.026}}.

    \bibitem{mienye2025deep}
    I.~D. Mienye, T.~G. Swart, Deep autoencoder neural networks: A comprehensive
    review and new perspectives, Archives of Computational Methods in Engineering
    32~(7) (2025) 3981--4000.
    \newblock \href {http://dx.doi.org/10.1007/s11831-025-10260-5}
    {\path{doi:10.1007/s11831-025-10260-5}}.

    \bibitem{sewak2020overview}
    M.~Sewak, S.~K. Sahay, H.~Rathore, An overview of deep learning architecture of
    deep neural networks and autoencoders, Journal of Computational and
    Theoretical Nanoscience 17~(1) (2020) 182--188.
    \newblock \href {http://dx.doi.org/10.1166/jctn.2020.8648}
    {\path{doi:10.1166/jctn.2020.8648}}.

    \bibitem{nawaz2024ensemble}
    A.~Nawaz, S.~S. Khan, A.~Ahmad, Ensemble of autoencoders for anomaly detection
    in biomedical data: A narrative review, {IEEE Access} 12 (2024) 17273--17289.
    \newblock \href {http://dx.doi.org/10.1109/ACCESS.2024.3360691}
    {\path{doi:10.1109/ACCESS.2024.3360691}}.

    \bibitem{Senthilcyber}
    R.~Senthil, S.~L. Wong, \href{https://arxiv.org/abs/2510.21837}{Quantum
        autoencoders for anomaly detection in cybersecurity} (2025).
    \newblock \href {http://arxiv.org/abs/2510.21837} {\path{arXiv:2510.21837}}.
    \newline\urlprefix\url{https://arxiv.org/abs/2510.21837}

    \bibitem{articleDaoud}
    A.~Daoud, W.~Khedr, O.~Elkomy, K.~Hosny,
    \href{https://www.ijci.zu.edu.eg/index.php/ijci/article/view/82}{New
        autoencoder-based method for efficient anomaly detection in {ECG}
        bio-signals}, International Journal of Computers and Informatics (Zagazig
    University) 5 (2024) 1--12.
    \newline\urlprefix\url{https://www.ijci.zu.edu.eg/index.php/ijci/article/view/82}

    \bibitem{sun2024unsupervised}
    X.~Sun, S.~Guo, S.~Liu, J.~Guo, B.~Du, Unsupervised anomaly detection of
    nuclear power plants under noise background based on convolutional
    adversarial autoencoder combining self-attention mechanism, Nuclear
    Engineering and Design 428 (2024) 113493.
    \newblock \href {http://dx.doi.org/10.1016/j.nucengdes.2024.113493}
    {\path{doi:10.1016/j.nucengdes.2024.113493}}.

    \bibitem{racheal2022evaluation}
    S.~Racheal, Y.~Liu, A.~Ayodeji, Evaluation of optimized machine learning models
    for nuclear reactor accident prediction, Progress in Nuclear Energy 149
    (2022) 104263.
    \newblock \href {http://dx.doi.org/10.1016/j.pnucene.2022.104263}
    {\path{doi:10.1016/j.pnucene.2022.104263}}.

    \bibitem{Moloko2023prediction}
    L.~E. Moloko, P.~M. Bokov, X.~Wu, K.~N. Ivanov, Prediction and uncertainty
    quantification of {SAFARI-1} axial neutron flux profiles with neural
    networks, Annals of Nuclear Energy 188 (2023) 109813.
    \newblock \href {http://dx.doi.org//10.1016/j.anucene.2023.109813}
    {\path{doi:/10.1016/j.anucene.2023.109813}}.

    \bibitem{IAEA2022}
    {International Atomic Energy Agency}, {Research Reactor Benchmarking Database:
        Facility Specification and Experimental Data (Revision)}, Tech. Rep. No. 480
    (Rev.1), {International Atomic Energy Agency}, Vienna (2022).

    \bibitem{Prinsloo2017recent}
    R.~H. Prinsloo, F.~A. Van~Heerden, D.~Botes, R.~Mudau, Recent developments of
    the {OSCAR} calculational system, as applied to selected examples from {IAEA}
    research reactor benchmarks, in: {18th conference of the International Group
        on Research Reactors and IAEA workshop on safety reassessment of research
        reactors in light of the lessons learned from the Fukushima Daiichi accident
        (J7-TR-54790). Book of Abstracts}, no. {INIS-AU--0091} in International
    Nuclear Information System (INIS), 2017, pp. 15--15.

    \bibitem{Pandas}
    \href{https://pandas.pydata.org/}{{Pandas - Python Data Analysis Library}}.
    \newline\urlprefix\url{https://pandas.pydata.org/}

    \bibitem{bressert2012scipy}
    E.~Bressert, {SciPy and NumPy: An Overview for Developers}, O'Reilly Media,
    2012.

    \bibitem{scikit}
    \href{https://scikit-learn.org/stable/}{{Scikit-learn: Machine Learning in
            Python — Scikit-Learn 1.8.0 documentation}}.
    \newline\urlprefix\url{https://scikit-learn.org/stable/}

    \bibitem{dawood2013correction}
    D.~Mohammad, J.~Xiaoyi, K.~Sch{\"a}fers, Correction Techniques in Emission
    Tomography, {CRC} Press, Taylor \& Francis Group, 2012, version Date:
    20120117.

    \bibitem{Demir2022comparison}
    S.~Demir, Comparison of normality tests in terms of sample sizes under
    different skewness and kurtosis coefficients, International Journal of
    Assessment Tools in Education 9 (2022) 397--409.
    \newblock \href {http://dx.doi.org/10.21449/ijate.1101295}
    {\path{doi:10.21449/ijate.1101295}}.

    \bibitem{Deng2020}
    D.~Deng, {DBSCAN} clustering algorithm based on density, in: 2020 7th
    International Forum on Electrical Engineering and Automation {(IFEEA)}, 2020,
    pp. 949--953.
    \newblock \href {http://dx.doi.org/10.1109/IFEEA51475.2020.00199}
    {\path{doi:10.1109/IFEEA51475.2020.00199}}.

    \bibitem{hahsler2019dbscan}
    M.~Hahsler, M.~Piekenbrock, D.~Doran, {DBSCAN: Fast Density-Based Clustering
        with R}, Journal of Statistical Software 91~(1) (2019) 1--30.
    \newblock \href {http://dx.doi.org/10.18637/jss.v091.i01}
    {\path{doi:10.18637/jss.v091.i01}}.

    \bibitem{wang2014robust}
    H.~Wang, F.~Nie, H.~Huang,
    \href{https://proceedings.mlr.press/v32/wangj14.html}{Robust distance metric
        learning via simultaneous l1-norm minimization and maximization}, in: E.~P.
    Xing, T.~Jebara (Eds.), Proceedings of the 31st International Conference on
    Machine Learning, Vol.~32 of Proceedings of Machine Learning Research, PMLR,
    Bejing, China, 2014, pp. 1836--1844.
    \newline\urlprefix\url{https://proceedings.mlr.press/v32/wangj14.html}

    \bibitem{blogdbscan}
    R.~Bedre, \href{https://www.reneshbedre.com/blog/dbscan-python.html}{{DBSCAN
            Clustering in Python: A Complete Guide}}, accessed: 2026-06-18 (2024).
    \newline\urlprefix\url{https://www.reneshbedre.com/blog/dbscan-python.html}

    \bibitem{Keras}
    \href{https://www.tensorflow.org/guide/keras}{{Keras: The high-level API for
            TensorFlow}}.
    \newline\urlprefix\url{https://www.tensorflow.org/guide/keras}

    \bibitem{Andonie2019hyperparameter}
    R.~Andonie, Hyperparameter optimization in learning systems, Journal of
    Membrane Computing 1~(4) (2019) 279--291.
    \newblock \href {http://dx.doi.org/10.1007/s41965-019-00023-0}
    {\path{doi:10.1007/s41965-019-00023-0}}.

    \bibitem{loshchilov2017fixing}
    I.~Loshchilov, F.~Hutter, \href{https://arxiv.org/abs/1711.05101v1}{Fixing
        weight decay regularization in adam} (2017).
    \newblock \href {http://arxiv.org/abs/1711.05101v1}
    {\path{arXiv:1711.05101v1}}.
    \newline\urlprefix\url{https://arxiv.org/abs/1711.05101v1}

    \bibitem{komadina2024comparing}
    A.~Komadina, M.~Martini{\'c}, S.~Gro{\v{s}}, {\v{Z}eljka}.~Mihajlovi{\'c},
    Comparing threshold selection methods for network anomaly detection, {IEEE
        Access} 12 (2024) 124943--124973.
    \newblock \href {http://dx.doi.org/10.1109/ACCESS.2024.3452168}
    {\path{doi:10.1109/ACCESS.2024.3452168}}.

    \bibitem{hossin2015review}
    M.~Hossin, M.~N. Sulaiman, A review on evaluation metrics for data
    classification evaluations, International Journal of Data Mining \& Knowledge
    Management Process 5~(2) (2015) 1.
    \newblock \href {http://dx.doi.org/10.5121/ijdkp.2015.5201}
    {\path{doi:10.5121/ijdkp.2015.5201}}.

\end{thebibliography}

\end{document}